\documentclass[11pt,a4paper]{article}

\usepackage[T1]{fontenc}
\usepackage[utf8]{inputenc}
\usepackage{lmodern}
\usepackage{microtype}
\usepackage{geometry}
\usepackage{amsmath,amssymb,amsthm,mathtools,bm}
\usepackage{booktabs,tabularx,array}
\usepackage{float}
\usepackage{enumitem}
\usepackage{xcolor}
\usepackage{hyperref}
\hypersetup{
  colorlinks=true,
  linkcolor=black,
  citecolor=black,
  urlcolor=black,
  pdftitle={Dynamical and Stochastic Limits of Non-Semisimple Integrable Spin Chains},
  pdfauthor={Anton Pribytok}
} 
\usepackage{bookmark}

\numberwithin{equation}{section}

\theoremstyle{definition}

\newcommand{\tx}[1]{\text {#1}}
\newcommand{\cl}[1]{\mathcal {#1}}

\newcommand{\bb}[1]{\mathbb {#1}}

\newcommand{\mb}[1]{\mathbf {#1}}

\newcommand{\id}{\mathbf 1}
\newcommand{\cH}{\mathcal H}
\newcommand{\cN}{\mathcal N}
\newcommand{\cL}{\mathcal L}
\newcommand{\cR}{\mathcal R}
\newcommand{\cM}{\mathcal M}
\newcommand{\End}{\operatorname{End}}
\newcommand{\Tr}{\operatorname{Tr}}
\newcommand{\spec}{\operatorname{spec}}
\newcommand{\rank}{\operatorname{rank}}
\newcommand{\ket}[1]{\lvert #1\rangle}
\newcommand{\bra}[1]{\langle #1\rvert}
\newcommand{\E}{E_{14}}

\title{\textbf{Non-Semisimple Integrability and Dynamical Limits}}

\author{
	Anton Pribytok$^{1,2,3}$\\[2mm]
	\small $^{1}$Beijing Institute of Mathematical Sciences and Applications (BIMSA),\\
	\small Huairou District, Beijing 101408, China\\[1mm]
	\small $^{2}$Yau Mathematical Sciences Center (YMSC), Tsinghua University,\\
	\small Haidian District, Beijing 100084, China\\[1mm]
	\small $^{3}$Faculty of Mathematics, National Research University Higher School of Economics,\\
	\small Usacheva str. 6, 119048 Moscow, Russia\\[1mm]
	\small \texttt{antonspribitoks@bimsa.cn}
}

\date{}

\begin{document}
\maketitle 

\begin{abstract}
	We investigate Jordan structure and dynamical limits in six families of
	integrable spin-$\frac12$ chains obtained from the boost automorphism method.
	Local operator algebras connect nilpotent evolution on finite periodic
	chains to central defects in spectral additivity, polynomial
	unipotent $R$-matrices and constant braid limits. For conservative endpoint
	erosion, we determine the characteristic and minimal polynomials and
	show that the Jordan partition can change while both polynomials remain
	unchanged. We also identify the deletion limits that distinguish the parity obstruction to removing signs by diagonal similarity in Class 5 from the parity nilpotent dependence in Class 6. Matching expansions and fragmentation
	recursions yield exact amplitudes and survival probabilities for
	processes with killing. In addition, we establish criteria excluding
	conservative Markov realizations and determine which conservative
	completions of specified reaction densities admit regular
	$R$-matrices of difference form.
\end{abstract}

\newpage 

\tableofcontents 

\newpage 

\section{Introduction}
\label{sec:introduction}

Quantum integrability provides a common framework for quantum spin chains,
vertex models and stochastic systems with integrable generators. The
Yang--Baxter equation (YBE) \cite{Yang1967,Baxter1972} connects local interactions to commuting transfer
matrices and conserved charges, but neither commutativity nor integrability
implies hermiticity, diagonalizability or probability conservation. Parametric and spectral limits can further isolate different sectors of the local algebra, leading to nilpotent generators, constant braid operators or stochastic representatives. These
structures must therefore be analysed separately. Our aim is to determine
how the local Yang-Baxter algebra controls exact finite chain dynamics and which algebraic and probabilistic properties survive its distinguished limits, with emphasis on minimal polynomials, Jordan structure and exact propagators. 

Our starting point is the classification of regular analytic
$4\times4$ $R$-matrices of difference form \cite{deLeeuwPribytokRyan2019}. It contains fourteen families under the integrable identifications implemented there, including eight familiar vertex families and six additional integrable model classes. The present work does not repeat that classification. Instead,
we determine how these six novel families support nilpotent evolution, constant braid limits and put particular emphasis on their relation to stochastic dynamics.

For this purpose, we take $V=\mathbb C^2$ with basis $\{\ket0,\ket1\}$ and use the ordered
basis \newline $\{\ket{00},\ket{01},\ket{10},\ket{11}\}$ on $V\otimes V$.
For a regular difference form solution, the permutation $P$ and the
braid matrix $\check R=PR$ it conventionally provides 
\begin{equation}
	R(0)=P,
	\qquad
	\check R(0)=\id,
	\qquad
	h=\check R'(0),
	\qquad
	\cH^{(L)}=Q_2=\sum_{j=1}^{L}h_{j,j+1},
	\label{eq:intro-regularity}
\end{equation}
where site labels are periodic and $\check R$ satisfies the Yang-Baxter equation in braid form 
\begin{equation}
	\check R_{12}(u)\check R_{23}(u+v)\check R_{12}(v)
	=
	\check R_{23}(v)\check R_{12}(u+v)\check R_{23}(u).
	\label{eq:intro-braid-ybe}
\end{equation}
It also implies commutativity of the transfer matrices
$\tau(u)=\Tr_a[R_{aL}(u)\cdots R_{a1}(u)]$. Since $\tau(0)$ corresponds to the translation operator, the expansion 
\begin{equation}
	\log\!\bigl(\tau(u)\tau(0)^{-1}\bigr)
	=\sum_{r=2}^{\infty}\frac{u^{r-1}}{(r-1)!}Q_r,
	\qquad
	[Q_r,Q_s]=0
	\label{eq:intro-charges}
\end{equation}
generates the local conserved hierarchy in involution without any assumption of hermiticity or diagonalizability \cite{KulishSklyanin1982,deLeeuwEtAl2021}. The boost construction provides a direct suitable way from the density to this hierarchy. Following the lattice boost and ladder operator formalism \cite{Tetelman1982,LinksEtAl2001}, one can introduce
\begin{equation}
	\mathcal B=\sum_{j\in\mathbb Z}j\,h_{j,j+1},
	\qquad
	Q_{r+1}=[\mathcal B,Q_r],
	\qquad
	Q_3=-\sum_j[h_{j,j+1},h_{j+1,j+2}],
	\label{eq:intro-boost}
\end{equation}
for $r\geq2$, with compatible charge normalisations and scalar identity
terms being omitted from the recursion. The sign of $Q_3$ follows from the displayed choice
of $\mathcal B$. The weighted sum is formal on the infinite chain and one first needs to derive its local commutators and then to sum the resulting
densities periodically, rather than treating $\mathcal B$ as a finite periodic operator. Its adjoint action corresponds to derivation, whose formal
exponential gives the boost automorphism
\begin{equation}
	e^{s\operatorname{ad}_{\mathcal B}}Q_r
	=\sum_{n=0}^{\infty}\frac{s^n}{n!}Q_{r+n},
	\qquad
	\operatorname{ad}_{\mathcal B}X=[\mathcal B,X].
	\label{eq:intro-boost-automorphism}
\end{equation}
On the formal charge generating series, this action translates the
spectral parameter, which is not the local composition law $\check R(u)\check R(v)=\check R(u+v)$.

The classification method exploits this relation in the reverse
direction. Starting from a general two site density, one constructs
$Q_3$ and imposes $[Q_2,Q_3]=0$. For difference form models, this yields
polynomial constraints on the entries of $h$. A necessary local constraint is
the Reshetikhin condition
\begin{equation}
	[h_{12}+h_{23},[h_{12},h_{23}]]=X_{23}-X_{12},
	\qquad X\in\End(V\otimes V),
	\label{eq:intro-reshetikhin}
\end{equation}
where the two site operator $X$ accounts for a telescoping remainder
\cite[Eq.~(3.20)]{KulishSklyanin1982}.
Candidate densities are then used to reconstruct $R$ through its regular
expansion or the Sutherland type equations \cite{Sutherland1970,deLeeuwEtAl2021}. This construction also explains why the local density must be retained
alongside its periodic sum. A lattice coboundary satisfies $\sum_{j=1}^{L}(\delta A)_{j,j+1}=0$,
It therefore leaves $Q_2$ unchanged. Preservation of
$Q_2$ alone does not guarantee that the modified density admits a regular
$R$-matrix. Separately, a scalar normalisation
$\widetilde R(u)=f(u)R(u)$ with $f(0)=1$ preserves the Yang-Baxter
equation, but changes the density to $\widetilde h=h+f'(0)\id$. These distinctions matter both for singular limits and for probability conservation. 

Our analysis proceeds from the local algebra to finite periodic dynamics induced by six classes of models in \cite{deLeeuwPribytokRyan2019}. We separate semisimple and nilpotent components, determine minimal
polynomials and compare the underlying Yang-Baxter family with the time exponential.
In the occupation basis, one can denote $\mathcal F_m$ to be the subspace with at most $m$ occupied sites. Strict lowering $\mathcal H^{(L)}\mathcal F_m\subseteq\mathcal F_{m-1}$ implies $\bigl(\mathcal H^{(L)}\bigr)^{L+1}=0$. It is established that when diagonal or exchange terms remain, lowering operators can instead generate Jordan chains. A block of size $d$ at eigenvalue $\lambda$
contributes $t^j e^{\lambda t}$ with $0\leq j\leq d-1$, so the minimal polynomial directly constrains the finite time dynamics. 

We distinguish parametric limits at fixed $u$, normalized limits as $u\to\pm\infty$ and multiple scaling limits of free parameters and $u$. These operations can retain different semisimple and nilpotent sectors and need not commute. A limiting operator is a constant braid solution, if and only if
\begin{equation}
	B_{12}B_{23}B_{12}=B_{23}B_{12}B_{23}.
\end{equation}
Important to note that neither idempotency, nor rank reduction at a particular spectral value is sufficient. In particular, fusion projectors and constant braid
projectors constitute distinct notions. 

In general stochasticity also imposes additional conditions. In the column convention, a conservative Markov generator has nonnegative off-diagonal entries and zero column sums \cite{AlcarazDrozHenkelRittenberg1994}. A nonzero
nilpotent operator cannot satisfy these conditions, while a nontrivial Jordan block at zero is incompatible with a bounded stochastic semigroup. At the same time Jordan blocks at strictly negative eigenvalues remain compatible with exponential relaxation. We therefore distinguish conservative dynamics, positive weighted evolution and processes with killing. A global basis
change that makes amplitudes nonnegative need not define a local stochastic transformation of the $R$-matrix, as well as certain configuration dependent
escape terms define a new density, whose integrability yet must be checked independently. 

The article proceeds through the six classes, Classes 1 and 2 establish
the role of periodic coboundaries, nilpotent deletion and central obstructions to spectral additivity. Then we proceed to Class 3m, which develops a conservative endpoint erosion and determines the maximal Jordan block size $L-2k+1$ at eigenvalue $-k\gamma$, where $k$ counts occupied clusters
and $\gamma$ is the sum of the endpoint rates. Next, the Class 4 relates a central Jordan defect to reductions to Class 3, mixed braid limits and exact matching amplitudes. Class 5 separates exchange from signed
endpoint deletion and identifies the parity obstruction to a diagonal change of basis with nonnegative amplitudes. Then we investigate Class 6 that develops
fragmentation after exchange is removed, with nilpotency indices for even and odd $L$ at nonzero lowering coupling and generating functions for complete and partial deletion histories. We also provide analysis of pair deletion limits that connect the matching sector to conservative
reaction models for which the stochastic completion is distinct from the initial integrable density. 

\section{Class 1: Exact exponential flow and deletion dynamics}

\subsection{Local algebra}

The Class-1 has the corresponding density \cite{deLeeuwPribytokRyan2019}
\begin{equation}
 h_1=
 \begin{pmatrix}
 0&a_1&a_2&0\\
 0&a_5&0&a_3\\
 0&0&-a_5&a_4\\
 0&0&0&0
 \end{pmatrix},
 \qquad
 a_1a_3-a_2a_4=0.
 \label{eq:h1}
\end{equation}
From the direct multiplication one can obtain the off-shell relation in the form 
\begin{equation}
 h_1^3-a_5^2h_1
 =a_5(a_1a_3-a_2a_4)\E,
 \qquad
 \E=\ket{00}\bra{11} \,,
 \label{eq:h1-cubic-offshell}
\end{equation}
which on the integrable locus of Class 1 provides $ h_1^3=a_5^2h_1$\label{eq:h1-cubic} and for $a_5\neq0$ the minimal polynomial generically has the three distinct roots $0,\pm a_5$.  The local density is then diagonalisable even though the periodic many-body Hamiltonian appears nilpotent. To derive the class 1 condition and note its algebraic implications, we can fix the basis of single site operators
\begin{equation}
	e=\ket0\bra1,
	\qquad
	n=\ket1\bra1,
	\qquad
	\bar n=\ket0\bra0,
	\qquad
	z=\bar n-n.
\end{equation}
Then the density directly can be decomposed as 
\begin{equation}
	h_1=\mathfrak d_1+\mathfrak x_1+\mathfrak y_1,
	\label{eq:h1-decomposition}
\end{equation}
\begin{equation} 
	\mathfrak d_1=\frac{a_5}{2}(z\otimes\id-\id\otimes z), \quad 
	\mathfrak x_1=(a_1\bar n+a_4 n)\otimes e, \quad 
	\mathfrak y_1=e\otimes(a_2\bar n+a_3 n).
	\label{eq:DecompositionOperators_1}
\end{equation}
where the operators obey 
\begin{equation}
	\mathfrak x_1^2=\mathfrak y_1^2=0,
	\qquad
	[\mathfrak d_1,\mathfrak x_1]=-a_5\mathfrak x_1,
	\qquad
	[\mathfrak d_1,\mathfrak y_1]=a_5\mathfrak y_1,
	\label{eq:h1-solvable}
\end{equation}
\begin{equation}
	[\mathfrak x_1,\mathfrak y_1]
	=(a_1a_3-a_2a_4)\E.
	\label{eq:h1-commutator}
\end{equation} 
For clarity one can also introduce the two-site occupation operator $n_{12}=n\otimes\id+\id\otimes n$, which together with \([n,e]=-e\) leads to
\begin{equation}
	[n_{12},\mathfrak x_1]=-\mathfrak x_1,
	\qquad
	[n_{12},\mathfrak y_1]=-\mathfrak y_1,
	\qquad
	[n_{12},\E]=-2\E.
	\label{eq:h1-occupation-grading}
\end{equation} 
Hence $\mathfrak x_1$ and $\mathfrak y_1$ are nilpotent operators of occupation degree $-1$ and the class 1 constraint \ref{eq:h1} is exactly the condition that the two occupation lowering operators commute. In other means it identifies the amplitudes of two ordered paths from $\ket{11}$ to $\ket{00}$.


\subsection{Spectral additivity and resolution} Another important property that arises for class 1 and appears important for further derivation of the underlying quantum algebras is the spectral, which generally fails for the class 2. By using \eqref{eq:h1-cubic}, it can be noticed that it exhibits truncation of the corresponding matrix exponential 
\begin{equation}\textbf{}
 e^{u h_1}
 =\id+\frac{\sinh(a_5u)}{a_5}h_1
 +\frac{\cosh(a_5u)-1}{a_5^2}h_1^2 \,.
 \label{eq:h1-exp-polynomial}
\end{equation}
Substituting the powers of $h_1$ reproduces the $R$-matrix of the initial model
\begin{equation}
\check R_1(u)=P R_1(u)=e^{u h_1} \,, 
\label{eq:R1-exponential}
\end{equation}
which provides 
\begin{equation}
\check R_1(u)\check R_1(v)=\check R_1(u+v),
\qquad
\check R_1(u)^{-1}=\check R_1(-u).
\label{eq:R1-group}
\end{equation}
In class 1 the spectral parameter therefore has an exact additive interpretation at the two-site level. In fact, this property is stronger than regularity of the initial proposition. From the algebra \eqref{eq:h1-solvable} and \([\mathfrak x_1,\mathfrak y_1]=0\) one can also derive the ordered factorisation 
\begin{equation}
	\check R_1(u)
	=
	e^{u\mathfrak d_1}
	\left(
	\id+\frac{e^{a_5u}-1}{a_5}\,\mathfrak x_1
	\right)
	\left(
	\id+\frac{1-e^{-a_5u}}{a_5}\,\mathfrak y_1
	\right).
	\label{eq:R1-factorisation}
\end{equation}
Hence we can establish that integrable locus $a_1a_3=a_2a_4$ implies the equality of two ordered paths from $\ket{11}$ to $\ket{00}$ 
\begin{equation}
\ket{11}\xrightarrow{Y_1}a_3\ket{01}
\xrightarrow{X_1}a_1a_3\ket{00} 
\qquad
\ket{11}\xrightarrow{X_1}a_4\ket{10}
\xrightarrow{Y_1}a_2a_4\ket{00} \,. 
\end{equation} 
To make asymptotic limits transparent it is suitable to consider the following decomposition of $\check{R}$, provided $a_5\neq0$
\begin{equation}
	\check R_1(u)=e^{a_5u}\Pi_++e^{-a_5u}\Pi_-+\Pi_0 \,,
	\label{eq:R1-spectral}
\end{equation} 

\begin{equation}
	\Pi_+ =\frac{h_1(h_1+a_5\id)}{2a_5^2} \, \qquad
	\Pi_- =\frac{h_1(h_1-a_5\id)}{2a_5^2} \, \qquad
	\Pi_0 =\id-\frac{h_1^2}{a_5^2} \,.
\end{equation} 
which explicitly subject to 
\begin{equation}
\Pi_\alpha\Pi_\beta=\delta_{\alpha\beta}\Pi_\alpha,
\qquad
\Pi_++\Pi_-+\Pi_0=\id,
\qquad
h_1=a_5(\Pi_+-\Pi_-).
\end{equation}
In general $\Pi_\pm$ have rank one, whereas $\Pi_0$ has rank two.

\subsection{Periodicity and asymptotic limit} 
Now let us notice the implications of boundary conditions on the underlying algebra and charge hierarchy. It can be noticed that the diagonal component $\mathfrak d_1$ in \eqref{eq:DecompositionOperators_1} corresponds to a lattice coboundary, which up to sign convention is 
\begin{equation}
	\mathfrak d_1
	=
	g\otimes\id-\id\otimes g,
	\qquad
	g=\frac{a_5}{2}z.
\end{equation}
which telescopically vanishes on the periodic chain 
\begin{equation}
	\mathcal H_1^{(L)}
	=
	\sum_{j=1}^{L}(h_1)_{j,j+1},
	\quad
	L+1\equiv1 \,: \qquad 
	\sum_{j=1}^{L}(\mathfrak d_1)_{j,j+1}
	=
	\frac{a_5}{2}
	\sum_{j=1}^{L}(z_j-z_{j+1})
	=0 \,.
\end{equation}
Thus $a_5$ is nontrivial at the level of the local density and
its operator algebra, but drops at the level of the first charge on a periodic chain. In the case of an open chain the same term instead reduces to the boundary contribution. Hence the periodic Hamiltonian is independent of $a_5$ 
\begin{equation}
	\cH_1^{(L)}
	=\sum_{j=1}^{L}
	\Bigl[
	(a_1\bar n_{j-1}+a_4n_{j-1})e_j
	+e_j(a_2\bar n_{j+1}+a_3n_{j+1})
	\Bigr].
	\label{eq:H1-global}
\end{equation}
Hence every nonzero term in \eqref{eq:H1-global} changes one local state $\ket{1}\to\ket{0}$, and therefore
\begin{equation}
	[\cN,\cH_1^{(L)}]=-\cH_1^{(L)} \qquad \qquad \cN=\sum_{j=1}^{L}n_j \,,
	\label{eq:H1-grading}
\end{equation}
where $\cl{N}$ constitutes total occupation operator and it immediately follows 
\begin{equation}
	\bigl(\cH_1^{(L)}\bigr)^{L+1}=0.
	\label{eq:H1-global-nilpotent}
\end{equation}
For an initial state containing $m$ occupied sites, the evolution truncates already at order $m$:
\begin{equation}
	e^{t\cH_1^{(L)}}\ket{C_m}
	=\sum_{k=0}^{m}\frac{t^k}{k!}
	(\cH_1^{(L)})^k\ket{C_m}.
	\label{eq:H1-finite-evolution}
\end{equation}
Thus matrix elements are weighted sums over so called \textit{ordered deletion histories}. This gives a clear continuous time transfer interpretation, although this is not yet a conservative stochastic process. On the other hand, the constraint in \eqref{eq:h1} can be parametrised through rank one matrix as 
\begin{equation}
	\begin{pmatrix}a_1&a_2\\a_4&a_3\end{pmatrix}
	=\begin{pmatrix}x\\y\end{pmatrix}
	\begin{pmatrix}p&q\end{pmatrix}.
	\label{eq:h1-rankone}
\end{equation}
By considering $d=x \bar n+yn$, we can define 
\begin{equation}
	\cL=\sum_jd_{j-1}e_j,
	\qquad
	\cR=\sum_je_jd_{j+1}.
\end{equation} 
which provides explicit separation into commuting left- and right lowering operators (chiral flows) for the periodic dynamics 
\begin{equation}
	\cH_1^{(L)}=p\cL+q\cR,
	\qquad
	[\cL,\cR]=0 \,, 
	\label{eq:H1-chiral}
\end{equation}
\begin{equation}
	e^{t\cH_1^{(L)}}=e^{tp\cL}e^{tq\cR}.
\end{equation}
In the derived framework the limits $p=0$ or $q=0$ are strictly chiral. Choosing $d=\bar n$ would provide a boundary-erosion transition graph, whereas $d=n$ gives a facilitated deletion graph. So far this structure establishes the transition skeleton and weighted amplitudes, for the probability conservation one requires additional diagonal "escape" terms.

\paragraph{Nilpotency.} Since $a_5$ enters the periodic Hamiltonian only through the vanishing coboundary, the limit $a_5\to0$ can be viewed primarily as a contraction of the $R$-matrix rather than a change of the periodic $Q_2$. The limiting density becomes 
\begin{equation}
	\bb{N}_1=
	\begin{pmatrix}
		0&a_1&a_2&0\\
		0&0&0&a_3\\
		0&0&0&a_4\\
		0&0&0&0
	\end{pmatrix},
	\qquad a_1a_3=a_2a_4 \,, 
\end{equation}
which obeys 
\begin{equation}
	\bb{N}_1^2=(a_1a_3+a_2a_4)\E=2a_1a_3\E,
	\qquad
	\bb{N}_1^3=0.
	\label{eq:N1-powers}
\end{equation}
Hence the contracted braid matrix $ \check{R}_1 $ is an exact polynomial unipotent family
\begin{equation}
		\check R_1^{\rm nil}(u)
		=e^{u\bb{N}_1}
		=\id+u\bb{N}_1+\frac{u^2}{2}\bb{N}_1^2. 
	\label{eq:R1-unipotent}
\end{equation}
In the case of generic parameters with $\bb{N}_1^2\neq0$, the local Jordan type is $(3,1)$. Further parametric restrictions would produce square zero constraints of type $(2,2)$ or $(2,1,1)$. 

\paragraph{Large $ \mb u$ limits.} Assume $a_5>0$ and $u$ real, then \eqref{eq:R1-spectral} results in 
\begin{equation}
	\lim_{u\to+\infty}e^{-a_5u}\check R_1(u)=\Pi_+,
	\qquad
	\lim_{u\to-\infty}e^{a_5u}\check R_1(u)=\Pi_-.
	\label{eq:R1-projector-limits}
\end{equation}
These are rank one idempotents and direct substitution also gives the degenerate adjacent bond relations 
\begin{equation}
	(\Pi_\pm)_{12}(\Pi_\pm)_{23}(\Pi_\pm)_{12}=0,
	\qquad
	(\Pi_\pm)_{23}(\Pi_\pm)_{12}(\Pi_\pm)_{23}=0,
\end{equation}
so each projector $\Pi$ is a non-invertible constant braid solution. The order of limits also matters, since if $a_5\to0$ is taken first, then
\begin{equation}
	\lim_{u\to\infty}\frac{1}{u^2}\check R_1^{\rm nil}(u)
	=\frac12\bb{N}_1^2=a_1a_3\E,
	\label{eq:R1-squarezero-limit}
\end{equation}
which is square zero rather than idempotent and changes the semisimple spectral decomposition into a nilpotent Jordan type structure. The projective and nilpotent asymptotic regimes are therefore different boundary sectors of the same spectral family (parametric space decomposes into qualitatively different loci). 

There are nevertheless two weaker stochastic type interpretations. In the first case, if $\{a_{i}\}\ge0 \,, i=\overline{1,4}$, the matrix $\cH_1^{(L)}$ has nonnegative off-diagonal entries and $e^{t\cH_1^{(L)}}$ is a positive weighted "history" transfer matrix (with columns not normalized). The second possibility is that for any finite $L$ we choose $\lambda$ greater than or equal to the maximal column sum of $\cH_1^{(L)}$ and 
\begin{equation}
\cM_{1,\lambda}=\cH_1^{(L)}-\lambda\id
\end{equation}
has nonnegative off-diagonal entries and non-positive column sums.  It is a sub-Markov generator describing a \textit{killed process} by
\begin{equation}
e^{t\cM_{1,\lambda}}=e^{-\lambda t}e^{t\cH_1^{(L)}}.
\end{equation}
The scalar regularization preserves the integrable structure up to normalization, but this does not produce reaction process with conserved probability. Hence a conservative killing process can be built by adding system dependent diagonal escape rates to the same transition graph. Its natural physical role is a non-unitary weighted deletion dynamics or a killing process transfer system, rather than a conservative Markov chain.

\section{Class 2: Nilpotent central extension}

\subsection{Density, central extension and $\Delta_{2}$ invariant}

The Class 2 density \cite{deLeeuwPribytokRyan2019} arises in the form 
\begin{equation}
 h_2=
 \begin{pmatrix}
 0&a_2&a_3-a_2&a_5\\
 0&a_1&0&a_4\\
 0&0&-a_1&a_3-a_4\\
 0&0&0&0
 \end{pmatrix}.
 \label{eq:h2}
\end{equation}
and the braid matrix can be given as 
\begin{equation}
\check R_2(u)
=u\left[
\frac{a_1}{\sinh(a_1u)}\id+h_2
+\frac{\tanh(a_1u/2)}{a_1}h_2^2
\right] 
\label{eq:R2-source}
\end{equation}
\begin{equation}
	\mathcal{R}_2(u)
	=\frac{\sinh(a_1u)}{a_1u}\check R_2(u)
	=\id+\frac{\sinh(a_1u)}{a_1}h_2
	+\frac{\cosh(a_1u)-1}{a_1^2}h_2^2 \,, 
	\label{eq:R2-hat}
\end{equation}
which in the prescribed normalisation satisfies 
\begin{equation}
	\mathcal{R}_2(u)\mathcal{R}_2(-u)=\id,
	\qquad
	\det\mathcal{R}_2(u)=1.
	\label{eq:R2-inverse}
\end{equation} 
So no finite real value of $u$ produces a rank drop in the regular
normalisation. Similarly to Class 1, the Class 2 decomposition can be
given by
\begin{equation}
	h_2
	=
	\mathfrak d_2+\mathfrak x_2+\mathfrak y_2+\mathfrak z_2,
	\label{eq:h2-decomp}
\end{equation}
with
\begin{equation}
	\begin{aligned}
		\mathfrak d_2
		&=
		\frac{a_1}{2}(z\otimes\id-\id\otimes z),
		\qquad &
		\mathfrak y_2
		&=
		e\otimes\bigl((a_3-a_2)\bar n+a_4n\bigr),
		\\[1mm]
		\mathfrak x_2
		&=
		\bigl(a_2\bar n+(a_3-a_4)n\bigr)\otimes e,
		\qquad &
		\mathfrak z_2
		&=
		a_5 e\otimes e=a_5\E.
	\end{aligned}
\end{equation}
The nonzero commutators are
\begin{equation}
	[\mathfrak d_2,\mathfrak x_2]
	=
	-a_1\mathfrak x_2,
	\qquad
	[\mathfrak d_2,\mathfrak y_2]
	=
	a_1\mathfrak y_2,
	\qquad
	[\mathfrak x_2,\mathfrak y_2]
	=
	C_2\E,
	\label{eq:h2-algebra}
\end{equation}
where it is suitable to denote
\begin{equation}
	C_2=a_3(a_2-a_3+a_4).
\end{equation}
It can be further noted that the direct pair-lowering operator
$\mathfrak z_2$ is central, and one can define
\begin{equation}
	\Delta_2
	=
	a_1a_5-C_2
	=
	a_1a_5-a_2a_3+a_3^2-a_3a_4.
	\label{eq:Delta2}
\end{equation}
Here we can analogously see that the quantity $C_2$ is the difference
between the two ordered sequential paths from $\ket{11}$ to $\ket{00}$,
\begin{align}
	\mathfrak x_2\mathfrak y_2\ket{11}
	&=
	a_2a_4\ket{00},
	\\[1ex]
	\mathfrak y_2\mathfrak x_2\ket{11}
	&=
	(a_3-a_2)(a_3-a_4)\ket{00}.
\end{align}
Thus $\Delta_2$ compares the ordering asymmetry of the sequential
channels with the direct pair channel
$a_5\ket{00}\bra{11}$, weighted by the semisimple splitting $a_1$.
On the locus $\Delta_2=0$ it can be immediately noticed that
\begin{equation}
	[\mathfrak x_2,\mathfrak y_2]
	=
	a_1\mathfrak z_2\,,
	\label{eq:h2-central-extension}
\end{equation}
which is exactly the centrally extended analogue of the commuting
Class 1 lowering algebra.

\subsection{Jordan-Chevalley decomposition, additivity and deletion dynamics}
From the derivation of power constraints we can obtain 
\begin{equation}
	h_2^3-a_1^2h_2=-a_1\Delta_2\E,
	\qquad
	h_2^4=a_1^2h_2^2.
	\label{eq:h2-polynomial}
\end{equation}
Thus the minimal polynomial divides $x^2(x^2-a_1^2)$ and for $\{ a_1,\,\Delta_2 \}\neq0$, the generic local Jordan structure develops as 
\begin{equation}
	J_2(0)\oplus J_1(a_1)\oplus J_1(-a_1) \,, 
\end{equation}
whereas $\Delta_2=0$ the zero-eigenvalue block becomes semisimple and$h_2$ is generically diagonalisable. For $a_1\neq0$ the spectral projectors onto the two nonzero eigenspaces can be defined by
\begin{equation}
	\Pi_\pm
	=
	\frac{\pm1}{2a_1^3}
	h_2^2\bigl(h_2\pm a_1\id\bigr).
	\label{eq:h2-projectors}
\end{equation}
The nilpotent part that is supported on the generalized zero-eigenspace corresponds to 
\begin{equation}
	N_0
	=
	h_2P_0
	=
	h_2-\frac{h_2^3}{a_1^2}
	=
	\frac{\Delta_2}{a_1}\E \,, \quad \qquad N_{0}^{2} = 0 \,, \quad P_0 = \id-\frac{h_2^2}{a_1^2}.
	\label{eq:h2-N0}
\end{equation}
So $\Delta_2=0$ is precisely the locus on which the
nilpotent zero-eigenspace contribution $N_0$ disappears. Therefore the Class 2 Hamiltonian density can be given by the following decomposition 
\begin{equation}
	h_2
	=
	a_1(\Pi_+-\Pi_-)+N_0 \,, \qquad 	[\Pi_\pm,N_0]=0 \,. 
	\label{eq:h2-Jordan-decomp}
\end{equation}
\paragraph{Class 2 distinction.} By Using \eqref{eq:h2-polynomial}, the normalized braid matrix 
\eqref{eq:R2-hat} can be written as
\begin{equation}
		\mathcal R_2(u)
		=
		e^{u h_2}
		+
		\frac{\Delta_2}{a_1^2}
		\bigl(\sinh(a_1u)-a_1u\bigr)\E.
	\label{eq:R2-exp-correction}
\end{equation}
or equivalently by its Jordan type form 
\begin{equation}
	\mathcal R_2(u)
	=
	e^{a_1u}\Pi_+
	+
	e^{-a_1u}\Pi_-
	+
	P_0
	+
	\frac{\sinh(a_1u)}{a_1}N_0.
	\label{eq:R2-spectral-Jordan}
\end{equation}
From this formula one can note that it isolates the essential difference between Classes 1 and
2.  The Class 2 normalized braid matrix is not, in general, the
exponential generated by its logarithmic derivative and distinction from direct additivity comes from the central square zero operator $\E$. Indeed, by considering 
\begin{equation}
	F(u)=\frac{\sinh(a_1u)-a_1u}{a_1^2},
	\quad \qquad
	[h_2,\E]=0,
	\qquad 
	\E^2=0, 
\end{equation}
we can obtain 
\begin{equation}
	\mathcal R_2(u)\mathcal R_2(v)
	-
	\mathcal R_2(u+v)
	=
	\Delta_2
	\bigl[F(u)+F(v)-F(u+v)\bigr]\E.
	\label{eq:R2-semigroup-defect}
\end{equation}
Hence for generic spectral values it follows 
\begin{equation}
	\mathcal R_2(u)\mathcal R_2(v)
	=
	\mathcal R_2(u+v)
	\quad\Longleftrightarrow\quad
	\Delta_2=0.
\end{equation}
So on the $\Delta_2=0$ locus one therefore acquires the ordered factorisation
\begin{equation}
	\mathcal R_2(u)
	=
	e^{u\mathfrak d_2}
	\left(
	\id+\frac{e^{a_1u}-1}{a_1}\mathfrak x_2
	\right)
	\left(
	\id+\frac{1-e^{-a_1u}}{a_1}\mathfrak y_2
	\right)
	\left(
	\id+\frac{1-e^{-a_1u}}{a_1}\mathfrak z_2
	\right), 
	\label{eq:R2-factorization}
\end{equation}
where together with the centrality of $\mathfrak z_2$ we have used
\begin{equation}
	[\mathfrak d_2,\mathfrak x_2]
	=
	-a_1\mathfrak x_2,
	\qquad
	[\mathfrak d_2,\mathfrak y_2]
	=
	a_1\mathfrak y_2,
	\qquad
	[\mathfrak x_2,\mathfrak y_2]
	=
	a_1\mathfrak z_2 . 
\end{equation}
Thus $\Delta_2=0$ is simultaneously the locus of local diagonalisability, exact baxterisation and closure of the
lowering algebra in the centrally extended form.

Now to identify the main structure of dynamics, we will first analyse consequences of periodicity and nilpotency. As in Class 1, the semisimple component $\mathfrak d_2$ is a lattice coboundary 
\begin{equation}
	\mathfrak d_2
	=
	g_2\otimes\id-\id\otimes g_2,
	\qquad
	g_2=\frac{a_1}{2}z.
	\label{eq:d2-coboundary}
\end{equation}
The periodic Hamiltonian is therefore independent of $a_1$ and can be given in the following form 
\begin{equation}
	\cH_2^{(L)}
	=
	\sum_j\Bigl[
	\bigl(
	a_2\bar n_{j-1}+(a_3-a_4)n_{j-1}
	\bigr)e_j
	+
	e_j
	\bigl(
	(a_3-a_2)\bar n_{j+1}+a_4n_{j+1}
	\bigr)
	+a_5e_je_{j+1}
	\Bigr].
	\label{eq:H2-global}
\end{equation}
However now it becomes useful to separate the two occupation channels 
\begin{equation}
	\cH_2^{(L)}
	=
	\cH_2^{(-1)}+\cH_2^{(-2)},
\end{equation}
where $\cH_2^{(-1)}$ consists of the two single-site lowering channels,
while $\cH_2^{(-2)}$ is the direct adjacent-pair lowering channel. One then immediately finds 
\begin{equation}
	[\cN,\cH_2^{(-1)}]
	=
	-\cH_2^{(-1)},
	\qquad
	[\cN,\cH_2^{(-2)}]
	=
	-2\cH_2^{(-2)} \quad \qquad \cN=\sum_{j=1}^{L}n_j 
	\label{eq:H2-filtration}
\end{equation}
Every nonzero monomial in $\cH_2^{(L)}$ therefore lowers the total occupation number by at least one. More precisely, on a state containing $m$ occupied sites it corresponds to 
\begin{equation}
	\bigl(\cH_2^{(L)}\bigr)^{m+1}\ket{C_m}=0 \quad \tx{and} \quad e^{t\cH_2^{(L)}}\ket{C_m}
	=
	\sum_{k=0}^{m}
	\frac{t^k}{k!}
	\bigl(\cH_2^{(L)}\bigr)^k\ket{C_m} \,,
\end{equation}
which is a finite polynomial, whose matrix elements enumerate ordered
"histories" containing both sequential single-site deletion and direct
adjacent-pair deletion. Although the parameter $a_1$ itself does not alter this periodic charge, it
remains nontrivial in the local density, its operator algebra and the higher charges.

\subsection{Nilpotency and asymptotic limit} 

We will now consider two complementary limiting regimes of the Class 2. Since the semisimple parameter $a_1$ enters the first charge only through the telescopic coboundary $\mathfrak d_2$, its $a_1\to0$ limit changes the local integrability without modifying $\cH_2^{(L)}$ and reduces the
semisimple splitting into a nilpotent polynomial family. In contrast,
the large $u$ limit at fixed $a_1\neq0$ isolates the nonzero spectral sectors and produces constant braid operators. Again, these two operations need not commute and therefore lead to distinct asymptotic regimes.

In analogy with Class 1, the limiting local generator and its nilpotent structure are given by
\begin{equation}
	\bb{N}_2:=\left.h_2\right|_{a_1=0},
	\qquad
	\bb{N}_2^2=\Gamma_2\E,
	\qquad
	\bb{N}_2^3=0.
	\label{eq:N2}
\end{equation}
where
\begin{equation}
	\Gamma_2
	=
	a_2a_4+(a_3-a_2)(a_3-a_4).
	\label{eq:Gamma2}
\end{equation}
By exploiting  regulated ratios 
\begin{equation}
	\frac{\sinh(a_1u)}{a_1}\longrightarrow u,
	\qquad
	\frac{\cosh(a_1u)-1}{a_1^2}
	\longrightarrow
	\frac{u^2}{2},
\end{equation}
one obtains the exact unipotent family 
\begin{equation}
		\mathcal R_2(u)
		\xrightarrow{\,a_1\to0\,}
		\check R_2^{\rm nil}(u)
		=
		e^{u\bb{N}_2}
		=
		\id+u\bb{N}_2+\frac{u^2}{2}\bb{N}_2^2 .
	\label{eq:R2-nil}
\end{equation}
Here the distinction between the source and regular normalisations
disappears in the degeneration limit, since
$\sinh(a_1u)/(a_1u)\to1$. For generic $\Gamma_2\neq0$, the nilpotency rank (index) is three and the
local Jordan type appears to be $(3,1)$. However on the $\Gamma_2=0$  locus, one instead has $\bb{N}_2^2=0$, so the polynomial family truncates already at first order 
\begin{equation}
	\check R_2^{\rm nil}(u)
	=
	\id+u\bb{N}_2.
\end{equation}
As in Class 1, the degeneration $a_1\to0$ leaves the periodic Hamiltonian unchanged, since $\mathfrak d_2$ is telescopic.

\paragraph{Large $\mb{u}$ asymptotics.}
For $a_1>0$ and real $u$, the source normalisation \eqref{eq:R2-source} gives the constant braids 
\begin{equation}
	B_\pm
	:=
	\lim_{u\to\pm\infty}\frac{\check R_2(u)}{|u|}
	=
	\pm h_2+\frac{h_2^2}{a_1}
	=
	2a_1\Pi_\pm\pm N_0.
	\label{eq:B2-decomp}
\end{equation}
Unlike Class~1, the nilpotent zero eigenspace contribution survives and in \eqref{eq:R2-spectral-Jordan}, its coefficient
grows at the same exponential rate as that of the projector. Hence we get the explicit obstruction to idempotency 
\begin{equation}
	B_\pm^2-2a_1B_\pm=\mp2\Delta_2\E,
	\qquad
	B_\pm^3=2a_1B_\pm^2, 
	\label{eq:B2-quadratic-defect}
\end{equation}
 and $B_\pm/(2a_1)$ reduces to the rank one braid projector
$\Pi_\pm$ precisely on $\Delta_2=0$. Otherwise, the braid operator retains a nontrivial Jordan block at zero and no scalar normalisation can make it idempotent. On the other hand taking $a_1\to0$ first, instead recovers the square zero asymptotic regime as discussed for Class 1, which is now governed by $\Gamma_2$ 
\begin{equation}
	\lim_{u\to+\infty}
	\frac{\check R_2^{\rm nil}(u)}{u^2}
	=
	\frac12\bb{N}_2^2
	=
	\frac{\Gamma_2}{2}\E.
	\label{eq:R2-squarezero-limit}
\end{equation}
For $\Gamma_2\neq0$ this regime is purely nilpotent, opposite to the surviving semisimple sector of $B_\pm$. 

\paragraph{Relation to Class 1.} The relation between Classes 1 and 2 is most transparent at the level of
the local lowering algebra.  Removing the direct pair channel and requiring
the two sequential single particle channels to commute provides 
\begin{equation}
	a_5=0
	\;\Longrightarrow\;
	\mathfrak z_2=0,
	\qquad
	C_2=a_3(a_2-a_3+a_4)=0
	\;\Longrightarrow\;
	\Delta_2=0,
	\qquad
	[\mathfrak x_2,\mathfrak y_2]=0 .
	\label{eq:C2-Class1-reduction}
\end{equation}
The density then reduces to
$h_2=\mathfrak d_2+\mathfrak x_2+\mathfrak y_2$ with
\begin{equation}
	[\mathfrak d_2,\mathfrak x_2]=-a_1\mathfrak x_2,
	\qquad
	[\mathfrak d_2,\mathfrak y_2]=a_1\mathfrak y_2,
	\qquad
	[\mathfrak x_2,\mathfrak y_2]=0,
	\label{eq:C2-Class1-algebra}
\end{equation}
which after relabelling of the couplings becomes precisely the same
solvable lowering algebra as in Class 1 and therefore admits analogous exponential factorisation. The new Class 2 structure appears exactly when the central pair channel
is retained. On the exponential locus we have 
\begin{equation}
	\Delta_2=0
	\quad\Longleftrightarrow\quad
	C_2=a_1a_5
	\quad\Longleftrightarrow\quad
	[\mathfrak x_2,\mathfrak y_2]
	=a_1\mathfrak z_2,
	\qquad
	\mathfrak z_2=a_5\E ,
	\label{eq:C2-central-relation}
\end{equation}
and the noncommutativity of two ordered single particle lowering paths is compensated by the pair lowering channel.  Class 2 therefore extends the commuting Class 1 and the last is recovered when the central generator $\mathfrak z_2$ is removed. 

\paragraph{Stochastic status.} The nilpotency of periodic Class 2 Hamiltonian obstructs nontrivial conservative continuous time Markov representative. Indeed for 
\begin{equation}
	\cM=\cl{S}\cH_2^{(L)}\cl{S}^{-1}-\lambda\id,
	\qquad
	\cl{S}\in GL\bigl((\bb C^2)^{\otimes L}\bigr),
	\qquad
	\spec(\cM)=\{-\lambda\},
\end{equation}
where $\cl{S}$ denotes an arbitrary invertible global basis change.
Conservation requires $0\in\spec(\cM)$ and hence $\lambda=0$, while
nilpotency gives $\Tr\cM=0$.  At the same time in the column convention a nonzero conservative Markov generator must satisfy 
\begin{equation}
	\Tr\cM=-\sum_C r_{\rm out}(C)<0,
	\qquad
	r_{\rm out}(C)=\sum_{C'\neq C}\cM_{C'C},
\end{equation}
hence in the given equivalence class there is no conservative Markov process. However there exists a positive, but non-conservative transfer regime. For example, in the occupation basis all off-diagonal amplitudes of \eqref{eq:H2-global} are nonnegative on \begin{equation}
	a_3\ge a_2\ge0,
	\qquad
	a_3\ge a_4\ge0,
	\qquad
	a_5\ge0,
	\label{eq:H2-positive-cone}
\end{equation}

\newpage 
\section{Class 3: Erosion, Jordan strata and singular limits}
\label{sec:n36-c3}

The strictly lowering part of the density now coexists with a diagonal
interaction and whose periodic sum does not vanish.  Consequently, the occupation
filtration remains useful, but no longer forces the Hamiltonian itself to be
nilpotent.  Instead, it separates the diagonal spectrum from the transitions
that can join states with the same eigenvalue.  Class 3 becomes particularly
transparent in this context, its two site density admits a semisimple resolution, while its periodic conservative representative develops Jordan blocks with maximal sizes that can be determined for every eigenvalue.

We shall retain the occupation basis and single site operators introduced above, and keep $E_{ij}$ for the corresponding two site matrix units. Throughout the following four sections, periodic chains have $L\geq3$ and stochastic matrices are acting on
probability columns. The initial input densities and spectral families are those of
\cite{deLeeuwPribytokRyan2019}. Their constant two state limits are considered
as boundaries of these particular spectral families, the constant classification problem has addressed in \cite{Hietarinta1993}.

\subsection{Spectral resolution and the conservative representative}

Let us set $\lambda_L=2a_1-a_2$ and $\lambda_R=2a_1+a_2$., the useful pattern of the
Class~3 density is that the same combinations multiply both the diagonal
splitting and the corresponding lowering entries
\begin{equation}
	h_3=\begin{pmatrix}
		-a_1&\lambda_La_3&\lambda_Ra_3&0\\
		0&a_1-a_2&0&0\\
		0&0&a_1+a_2&0\\
		0&0&0&-a_1
	\end{pmatrix},\qquad
	\begin{aligned}
		P_L&=E_{22}+a_3E_{12},\\
		P_R&=E_{33}+a_3E_{13},\\
		P_0&=\id-P_L-P_R.
	\end{aligned}
	\label{eq:n36-c3-input}
\end{equation}
The unit-matrix product rule gives $P_\alpha P_\beta=
\delta_{\alpha\beta}P_\alpha$ and $P_0+P_L+P_R=\id$.  These are generally
oblique projectors, the mutual annihilation is an algebraic statement, not orthogonality in the Hermitian inner product. Their ranks are accordingly $(2,1,1)$ and they therefore provide the 
\begin{equation}
	\begin{aligned}
		h_3&=-a_1\id+\lambda_LP_L+\lambda_RP_R,\\
		\check R_3(u)&=e^{uh_3}
		=e^{-a_1u}\bigl(P_0+e^{\lambda_Lu}P_L+e^{\lambda_Ru}P_R\bigr).
	\end{aligned}
	\label{eq:n36-c3-spectral}
\end{equation}
At an eigenvalue collision, the relevant projectors combine into the
projector onto the common eigenspace, i.e. they do not develop a nilpotent part. Thus $h_3$ remains diagonalizable for all finite parameters. After the coboundary contribution cancels, the diagonal part of the
periodic Hamiltonian is $-a_1\sum_j z_jz_{j+1}$. However, the full
densities on overlapping bonds generally do not need to commute and hence be simultaneously diagonalizable. 

In fact, the same resolution also makes conservative normalization possible.  For
$a_3\neq0$, we take $g=\operatorname{diag}(-1/a_3,1)$ and $G=g\otimes g$.
The diagonal similarity changes the two lowering amplitudes from
$\lambda_La_3,\lambda_Ra_3$ to $-\lambda_L,-\lambda_R$, while a scalar shift
sets the two diagonal entries to zero (uniform). More explicit 
\begin{equation}
	M_3=G(h_3+a_1\id)G^{-1}
	=\begin{pmatrix}
		0&r_L&r_R&0\\
		0&-r_L&0&0\\
		0&0&-r_R&0\\
		0&0&0&0
	\end{pmatrix},\qquad
	r_L=-\lambda_L,\quad r_R=-\lambda_R.
	\label{eq:n36-c3-markov}
\end{equation}
Every column sums to zero, hence it becomes Markov representative precisely when $r_L,r_R\geq0$ or equivalently $-2a_1\geq|a_2|$ for real parameters. For real $a_3<0$, the matrix $g$ is positive diagonal.  For other values
of $a_3$, it should be regarded only as an algebraic change of basis.
These statements are distinct since the transformed matrix may be
stochastic even when the transformation itself is not a positive reparametrisation of the original occupation basis. The $a_3=0$ is a
singular limit of this gauge transformation, where $g$ ceases to be invertible.

Because the family in \eqref{eq:n36-c3-spectral} is an exact
exponential, its transformed and scalar-normalized version is also the
local time semigroup:
\begin{equation}
	K_3(t)=e^{tM_3}=
	\begin{pmatrix}
		1&1-e^{-r_Lt}&1-e^{-r_Rt}&0\\
		0&e^{-r_Lt}&0&0\\
		0&0&e^{-r_Rt}&0\\
		0&0&0&1
	\end{pmatrix},\qquad t\geq0.
	\label{eq:n36-c3-kernel}
\end{equation}
It is simultaneously a stochastic kernel and an additive spectral
Yang--Baxter family. Its two reactions are $01\to00$ with rate $r_L$
and $10\to00$ with rate $r_R$. On the chain, these processes remove the
left or right endpoint of an occupied cluster adjacent to a vacancy.
Particle--hole exchange maps the same dynamics to irreversible growth
from occupied neighbors, with the symmetric case corresponding to
one-dimensional Richardson growth \cite{Richardson1973} and unequal
rates producing an anisotropic variant. Since occupation is deleted or
created rather than transported, the process does not become ASEP in the same
variables. The reaction rules and their empty-interval treatment are
established in \cite[Eqs.~(2), (6) and (8)]{AlimohammadiKhorramiAghamohammadi2001}. Here they
provide a concrete setting in which to determine the full finite-chain
minimal polynomial.

\subsection{Minimal polynomial and momentum resonance}

Summing the conservative density gives
\begin{equation}
	\mathcal M_3^{(L)}
	=\sum_{j=1}^L(r_L\bar n_{j-1}+r_R\bar n_{j+1})(e_j-n_j),
	\qquad \gamma=r_L+r_R>0.
	\label{eq:n36-c3-global}
\end{equation}

For $r_L,r_R\geq0$ with $\gamma>0$ we can establish that the characteristic and minimal
polynomials of the periodic erosion generator are given by \label{prop:n36-c3-minimal}
\begin{align}
	\chi_{\mathcal M_3^{(L)}}(x)
	&=x^2\prod_{k=1}^{\lfloor L/2\rfloor}
	(x+k\gamma)^{2\binom L{2k}},
	\label{eq:n36-c3-characteristic}\\
	m_{\mathcal M_3^{(L)}}(x)
	&=x\prod_{k=1}^{\lfloor L/2\rfloor}
	(x+k\gamma)^{L-2k+1}.
	\label{eq:n36-c3-minimal}
\end{align}
In particular, the largest Jordan block at $-k\gamma$ has size $L-2k+1$ and formula determines these maximal sizes, not all block multiplicities.

It can be derived  by considering a nonuniform configuration with $k$ occupied clusters, which has exactly $2k$ domain walls. Choosing their positions among the $L$ bonds and fixing the occupation of one reference site would give $2 \, \binom{L}{2k}$ configurations. In an occupation ordered basis all off-diagonal transitions lower the occupation number, so the diagonal escape rates determine the characteristic polynomial. In particular, the two uniform configurations have zero escape rate. In order to determine the minimal polynomial, one would need to consider the corresponding graded sector with fixed cluster number $k$. In this sector the generator acquires the form 
\begin{equation}
	-k\gamma \cdot \id+T_k,
\end{equation}
where $T_k$ contains precisely the endpoint deletions that preserve the number of clusters. in this case, a configuration with $k$ occupied clusters contains between $k$ and $L-k$ occupied sites, since every occupied cluster and every
separating vacant cluster contains at least one site. Each action of $T_k$ lowers the occupation by one while preserving $k$ being fixed, hence 
\begin{equation}
	T_k^{L-2k+1}=0.
\end{equation}
Clearly this bound is sharp and now we take $k$ separating vacancies, $k-1$ occupied
singletons (isolated occupied site with vacancies from both sides), and one occupied cluster of length $L-2k+1$. The long cluster can undergo $L-2k$ successive endpoint deletions before becoming a singleton, while the cluster number remains equal to $k$. At each step the sum of the left and right deletion amplitudes is $\gamma=r_L+r_R$, hence the total amplitude of these histories 
\begin{equation}
	\gamma^{L-2k}\neq0 \qquad \tx{and} \qquad  T_k^{L-2k}\neq0 \,, 
\end{equation}
with the $L-2k=0$ case representing that the corresponding graded sector is nonzero.

Finally, sectors with different cluster numbers carry distinct diagonal eigenvalues $-k\gamma$. Their off-diagonal "couplings" connect sectors with
different spectra and can be removed by block similarity, due to the fact that corresponding Sylvester maps are invertible. Important that these couplings can change the embedding of the generalized eigenspaces, but not their Jordan block sizes. The $k=0$ sector is spanned by the two uniform configurations is semisimple. Combining the maximal nilpotency indices of the fixed $k$ sectors gives \eqref{eq:n36-c3-minimal}.

From the minimal polynomial data we can also fix the possible time dependence without requiring a
complete Jordan basis.  Every propagator matrix element is a finite sum of
stationary terms and terms $t^j e^{-k\gamma t}$ with $0\leq j\leq L-2k$. These polynomial prefactors do not modify the eigenvalues. They arise from
the nilpotent part of the Jordan decomposition within a fixed primary
component. For a strictly negative eigenvalue $-k\gamma$, the resulting
terms $t^j e^{-k\gamma t}$ remain exponentially damped at long times, hence nontrivial Jordan blocks at negative eigenvalues are fully
compatible with a conservative generator.

The one cluster primary component admits a more explicit resolution and
shows that block multiplicities can change while the eigenvalues remain
fixed.  For instance, let us consider $\ket{j,m}$ be the configuration with one occupied cluster of
length $1\leq m<L$, starting at $j$. For this purpose, one can introduce the Fourier sums
\begin{equation}
	v_m(p)=\sum_{j=1}^L e^{-ipj}\ket{j,m},\qquad
	p=\frac{2\pi\ell}{L},\qquad q(p)=r_R+r_Le^{ip}.
	\label{eq:n36-c3-fourier}
\end{equation}
A right end deletion leaves the starting site unchanged, whereas a left end moves it from $j$ to $j+1$.  This shift supplies the phase $e^{ip}$,
and for $m\geq2$ provides 
\begin{equation}
	(\mathcal M_3^{(L)}+\gamma\id)v_m(p)=q(p)v_{m-1}(p).
	\label{eq:n36-c3-momentum}
\end{equation}
For a singleton, the only transition is to the vacuum.  Its Fourier sum
vanishes when $p\neq0$, so $v_1(p)$ is then an eigenvector at $-\gamma$.
At $p=0$, the vacuum coupling can be removed explicitly and by putting $w_m=v_m(0)-L\ket{0^L}$ it yields
\begin{equation}
	(\mathcal M_3^{(L)}+\gamma\id)w_1=0,\qquad
	(\mathcal M_3^{(L)}+\gamma\id)w_m=\gamma w_{m-1}\quad(m\geq2).
	\label{eq:n36-c3-zero-momentum-lift}
\end{equation}
This construction realizes the zero momentum chain ($p=0$) inside the generalized eigenspace at $-\gamma$, while the vacuum itself corresponds to the  zero eigenvalue sector.

If $q(p)\neq0$, the vectors at that momentum form a single block of size
$L-1$.  For nonnegative rates with $\gamma>0$, $q(p)$ can vanish only when
$r_L=r_R>0$ and $p=\pi$, which is an allowed momentum precisely for even
$L$. From here the primary Jordan structure at $-\gamma$ becomes immediately 
\begin{equation}
	\begin{cases}
		\displaystyle\bigoplus_{\ell=1}^{L}J_{L-1}(-\gamma),
		&L\text{ odd or }r_L\neq r_R,\\[1mm]
		\displaystyle\bigoplus_{\ell=1}^{L-1}J_{L-1}(-\gamma)
		\ \oplus\!\bigoplus_{\ell=1}^{L-1}J_1(-\gamma),
		&L\text{ even and }r_L=r_R>0.
	\end{cases}
	\label{eq:n36-c3-primary-resonance}
\end{equation}
For even $L$ with $r_L=r_R>0$, the alternating Fourier mode
$p=\pi$ satisfies $q(\pi)=0$, so the left and right endpoint
contributions cancel. This splits the corresponding Jordan chain and
changes the Jordan partition at $-\gamma$, while leaving its algebraic
multiplicity $L(L-1)$ unchanged. The maximal block size also remains
$L-1$, since all remaining Fourier sectors with $q(p)\neq0$ retain
their full Jordan chains.

For comparison with a direct observable, a vacancy bounded occupied cluster
of length $m$ passes through the lengths $m,m-1,\ldots,1$ with the same
total rate $\gamma$ at every step.  Its extinction time is therefore the
sum of $m$ independent exponential waiting times, giving
\begin{equation}
	\mathbb P(T_m>t)=e^{-\gamma t}
	\sum_{j=0}^{m-1}\frac{(\gamma t)^j}{j!}.
	\label{eq:n36-c3-extinction}
\end{equation}
For $m=L-1$, this reproduces the maximal time dependence
$t^{L-2}e^{-\gamma t}$ associated with the largest Jordan block in the
$-\gamma$ eigenspace. It does not determine the generalized eigenspaces
at $-k\gamma$ with $k\ge2$, whose maximal Jordan block sizes follow from
the cluster number and occupation number from above.

\subsection{Constant boundaries and deletion sector}

The spectral resolution also specifies which part of the local algebra survives at large rapidity. So for real $\lambda_L,\lambda_R$, we put
$\lambda_0=0$, $\mu_+=\max\{\lambda_0,\lambda_L,\lambda_R\}$ and
$\mu_-=\min\{\lambda_0,\lambda_L,\lambda_R\}$.  Then
\begin{equation}
	\lim_{u\to\pm\infty}e^{(a_1-\mu_\pm)u}\check R_3(u)
	=\sum_{\lambda_\alpha=\mu_\pm}P_\alpha \,, 
	\label{eq:n36-c3-constant}
\end{equation}
where all equal leading  exponents must retain (including at eigenvalue
collisions).  The resulting sums are idempotent constant braid operators. For fixed parameters, the same conclusion is consistent with the
braid-form Yang--Baxter equation when $u$, $v$, and $u+v$ are taken
simultaneously to $+\infty$ or $-\infty$ with the same normalization.

In the conservative regime with $r_L,r_R>0$, the long-time local kernel
leaves $\ket{00}$ and $\ket{11}$ unchanged and sends both mixed
configurations $\ket{01}$ and $\ket{10}$ to $\ket{00}$. Equivalently, for occupation variables $s_1,s_2\in\{0,1\}$, the map is
\begin{equation}
	(s_1,s_2)\longmapsto
	(s_1s_2,s_1s_2),
\end{equation}
where the product is the ordinary multiplication of the binary occupation numbers. The corresponding operator is a rank-two projector. If $r_L=0$, the state $\ket{01}$ is also fixed, while if $r_R=0$, the state $\ket{10}$ is fixed. In either case the limiting projector has rank three. For Class 3, the spectral parameter and local time generate the same family through \eqref{eq:n36-c3-kernel}, but this identification does not hold for the generic Class 4 family.

A different boundary removes the diagonal splitting while retaining the lowering amplitudes. By taking $a_1=\varepsilon A$, $a_2=\varepsilon B$, $a_3=\eta/\varepsilon$ at fixed
$u$, we consider the scaling limit 
\begin{equation}
	\nu_3
	=
	\beta_L\bar n\otimes e+\beta_Re\otimes\bar n,
	\qquad
	\beta_L=\eta(2A-B),
	\qquad
	\beta_R=\eta(2A+B).
	\label{eq:n36-c3-nu3}
\end{equation}
\begin{equation}
	\check R_3(u)\longrightarrow\id+u\nu_3,
	\qquad
	\nu_3^2=0.
	\label{eq:n36-c3-scaling-limit}
\end{equation}
The local square vanishes because either deletion produces the two-site
vacuum on which both channels vanish. The periodic sum nevertheless acts
nontrivially on longer clusters through overlapping bonds. It annihilates
the fully occupied periodic chain, but starting from a configuration with a
single vacancy it can remove all remaining $L-1$ particles. More precisely, if
\begin{equation}
	\mathcal V_3=\sum_j(\nu_3)_{j,j+1},
\end{equation}
then
\begin{equation}
	\mathcal V_3^L=0,
	\qquad
	\bra{0^L}\mathcal V_3^{L-1}\ket{0\,1^{L-1}}
	=
	(\beta_L+\beta_R)^{L-1}.
	\label{eq:n36-c3-scaling-index}
\end{equation}
Hence the nilpotency index of $\mathcal V_3$ is $L$ whenever
$\beta_L+\beta_R\neq0$, even though the local operator $\nu_3$ has
nilpotency index two. On the special locus
$\beta_R=-\beta_L\neq0$, the two deletion amplitudes of an isolated
occupied site cancel, and the periodic operator reduces to
\begin{equation}
	\mathcal V_3
	=
	\beta_L\sum_j e_j(n_{j+1}-n_{j-1}).
\end{equation}
This is precisely the signed endpoint eroding limit that will be analysed in Class 5. The scaling limit therefore connects the conservative erosion representative to a distinct nilpotent dynamics in which isolated occupied sites cannot be removed.

\section{Class 4: Central defect with the matching sector}
\label{sec:n36-c4}

Class~4 supplements the triangular Ising density with a direct operator that
removes two adjacent particles. Its amplitude combines with the sequential
single particle amplitudes to determine a local Jordan defect. The same
quantity controls the departure of the spectral family from the time
exponential. A uniform product similarity identifies the part of parameter
space already governed by Class~3, so only the additional algebraic and
dynamical structures need to be developed here.

\subsection{Reduction to Class 3 through defect}

Write $a=a_1$ throughout this section. The density and its quadratic identity
are
\begin{equation}
	h_4=\begin{pmatrix}
		a&a_2&a_2&a_3\\
		0&-a&0&a_4\\
		0&0&-a&a_4\\
		0&0&0&a
	\end{pmatrix},\qquad
	C_4=aa_3+a_2a_4,\qquad h_4^2-a^2\id=2C_4\E.
	\label{eq:n36-c4-input}
\end{equation}
The two contributions to $C_4$ come from direct pair removal and the two
ordered single particle paths. In this normalization, $C_4$ measures the
quadratic defect. Its numerical value depends on the chosen representative,
but the vanishing of $h_4^2-a^2\id$ is invariant under similarity.

Since $\E^2=0$ and $h_4\E=\E h_4=a\E$, for $a\neq0$ one can separate
commuting semisimple and nilpotent parts by setting
\begin{equation}
	S_4=h_4-\frac{C_4}{a}\E,\qquad
	P_{4,\pm}=\frac12\left(\id\pm\frac{S_4}{a}\right).
	\label{eq:n36-c4-semisimple}
\end{equation}
The quadratic identity gives $S_4^2=a^2\id$. Thus $P_{4,+}$ and $P_{4,-}$
are mutually annihilating projectors of rank two. Moreover,
$P_{4,+}\E=\E P_{4,+}=\E$ and $P_{4,-}\E=\E P_{4,-}=0$. The
nilpotent part therefore acts entirely within the generalized eigenspace at
$+a$. When $C_4\neq0$, its rank is one and the local Jordan form is
$J_2(a)\oplus J_1(-a)\oplus J_1(-a)$. On $C_4=0$ the density is
semisimple. These conclusions require $a\neq0$, and the limit $a\to0$
will be taken separately in the original matrix.

The $R$-matrix supplied by the classification and the time exponential
have the respective factorizations
\begin{align}
	\check R_4(u)
	&=e^{uS_4}\left[\id+\frac{C_4}{2a^2}\sinh(2au)\E\right],
	\label{eq:n36-c4-spectral}\\
	e^{th_4}
	&=e^{tS_4}\left[\id+\frac{C_4}{a}t\E\right].
	\label{eq:n36-c4-time}
\end{align}
The second identity follows by exponentiating the commuting parts of
$h_4$. In the first, the coefficient required by the Yang--Baxter family
is hyperbolic rather than linear. Both matrices equal $\id$ at zero and
have derivative $h_4$. Their difference at the same argument begins at
order $u^3$, but the two families coincide identically only when $C_4=0$.
Thus agreement of the infinitesimal density does not identify rapidity with
local physical time.

To expose the relation with Class~3, use the single site shear
$g_s=\id-a_4e/(2a)$ and its product $G_s=g_s\otimes g_s$. Its inverse
is obtained by changing the sign of the $e$ term. Direct conjugation gives
\begin{equation}
	G_sh_4G_s^{-1}
	=a\,z\otimes z+A(\bar n\otimes e+e\otimes\bar n)
	+\frac{C_4}{a}\E,\qquad A=a_2+a_4.
	\label{eq:n36-c4-normal}
\end{equation}
The shear removes the entries for single particle removal from $11$ and
collects their effect into the pair amplitude. On $C_4=0$ and $A\neq0$,
the remaining density is precisely symmetric Class~3, with its diagonal
parameter replaced by $-a$. For real $a>0$, a further diagonal single
site similarity with ratio $2a/A$, followed by subtraction of $a\id$,
gives the erosion density with $r_L=r_R=2a$.

The corresponding product transformation on the full chain therefore
transfers the characteristic and minimal polynomials in
\eqref{eq:n36-c3-characteristic} and \eqref{eq:n36-c3-minimal} to the
shifted Class~4 Hamiltonian. Here $\gamma=4a$, and the momentum resonance
in \eqref{eq:n36-c3-primary-resonance} occurs when the periodic chain has
even length. This conclusion follows from an explicit similarity, not
from equality of eigenvalues alone.

The additional condition $A=0$ gives a different result. On $C_4=A=0$,
\eqref{eq:n36-c4-normal} is simply $a\,z\otimes z$. A Hadamard rotation
and the same scalar shift give $a(\sigma^x\otimes\sigma^x-\id)$.
For $a>0$, this is a conservative generator of
$00\leftrightarrow11$ and $01\leftrightarrow10$, each at rate $a$.
The overlapping bond terms commute, so the periodic generator is
semisimple. This pair flip process preserves occupation parity rather than
occupation number. In domain wall variables it is the image of independent
Glauber spin flips at infinite temperature \cite{Glauber1963}. The two occupation parity sectors of the original periodic spin chain are
realized in the auxiliary Ising description by periodic and antiperiodic
boundary conditions for the auxiliary spins, respectively. This branch is
diagonalizable and therefore does not inherit the nontrivial Jordan structure
of the erosion representative with $A\neq0$.

These two constructions exhaust the conservative possibilities within the
specified local transformation problem. For real parameters with $a>0$,
there exist $g\in GL(2,\mathbb R)$ and a real scalar $\lambda$ such that
$(g\otimes g)h_4(g\otimes g)^{-1}-\lambda\id$ is a conservative
Markov generator if and only if $C_4=0$. Indeed, column conservation implies
that $(x,y)^{\otimes2}$ is a nonzero left eigenvector of $h_4$, where
$(x,y)=(1,1)g$. If $x\neq0$, the first component of its eigenvalue
equation forces $\lambda=a$. If $x=0$, the vector is proportional to
$\bra{11}$ and the same conclusion follows. The only possible shift is
therefore $-a\id$. For $C_4\neq0$, it places a nontrivial Jordan block at
zero, which would produce an unbounded time semigroup, whereas finite conservative
Markov semigroup is bounded. Necessity follows, while the erosion and pair
flip representatives establish sufficiency. 

\subsection{The surviving nilpotent boundary and killed kernel}

When $C_4\neq0$, allowing loss of probability gives a different
interpretation. Keeping real $a>0$ and $A\neq0$, we can apply the same product similarities and set 
\begin{equation}
	r=2a,\qquad \kappa=\frac{4aC_4}{A^2}.
	\label{eq:n36-c4-kappa}
\end{equation}
After subtraction of $a\id$, the density becomes $M_{\rm er}+\kappa\E$,
where $M_{\rm er}$ is symmetric erosion with endpoint rate $r$.
For $\kappa\geq0$, the off diagonal entries are nonnegative, but the
$11$ column has positive sum $\kappa$. A scalar subtraction gives
\begin{equation}
	\mathcal Q_{4,\sigma}=M_{\rm er}+\kappa\E-\sigma\id,
	\qquad \sigma\geq\kappa.
	\label{eq:n36-c4-time-killing}
\end{equation}
This is the generator of a killed Markov process. In state $11$, pair
removal has rate $\kappa$ and killing has rate $\sigma-\kappa$.
In every other state the killing rate is $\sigma$, in addition to the
erosion transitions. The identities $M_{\rm er}\E=\E M_{\rm er}=0$
give its exact local time evolution,
\begin{equation}
	e^{t\mathcal Q_{4,\sigma}}
	=e^{-\sigma t}\bigl[K_{\rm er}(t)+\kappa t\E\bigr].
	\label{eq:n36-c4-time-kernel}
\end{equation}
Here $K_{\rm er}$ is \eqref{eq:n36-c3-kernel} with both rates equal to
$r$. The linear pair term is the surviving Jordan contribution. 

At the level of the spectral family (Yang-Baxter), the corresponding sub-Markov
normalization corresponds to 
\begin{equation}
	B_\kappa(u)=
	\frac{K_{\rm er}(u)+(\kappa/r)\sinh(ru)\E}
	{1+(\kappa/r)\sinh(ru)},
	\qquad
	u\geq0.
	\label{eq:n36-c4-killed}
\end{equation}
By Introducing
\begin{equation}
	c(u)=\frac{\kappa}{r}\sinh(ru),
\end{equation}
the column sums become 
\begin{equation}
	\frac{1}{1+c(u)}
	\bigl(1,1,1,1+c(u)\bigr).
\end{equation}
For $\kappa\geq0$ all matrix elements are nonnegative and every column
sum does not exceed one. The missing probability can therefore be
interpreted as killing, or equivalently recovered by adjoining an
absorbing cemetery state. The common
scalar denominator preserves the Yang--Baxter equation, and
$B_\kappa'(0)=\mathcal Q_{4,\kappa}$. Nevertheless, for $\kappa\neq0$
this family is not $e^{u\mathcal Q_{4,\kappa}}$. The distinction arises both in the pair coefficient and in the scalar normalization. Equations \eqref{eq:n36-c4-time-kernel} and
\eqref{eq:n36-c4-killed} therefore describe a time semigroup and a
Yang--Baxter spectral family that share the same first derivative at the
regular point but differ at higher orders.

The local decomposition also fixes the asymptotic limits. Since
$e^{uS_4}\E=e^{au}\E$, the pair term in
\eqref{eq:n36-c4-spectral} is
$C_4(e^{3au}-e^{-au})\E/(4a^2)$. For fixed parameters and $a>0$ 
\begin{equation}
	\begin{array}{@{}l@{\qquad}l@{}}
		\displaystyle
		\lim_{u\to+\infty}e^{-3au}\check R_4(u)
		=\frac{C_4}{4a^2}\E
		&
		(C_4\neq0),
		\\[2mm]
		\displaystyle
		\lim_{u\to+\infty}e^{-au}\check R_4(u)
		=P_{4,+}
		&
		(C_4=0),
		\\[2mm]
		\displaystyle
		\lim_{u\to-\infty}e^{au}\check R_4(u)
		=b_{4,-}
		&
		b_{4,-}=P_{4,-}-\frac{C_4}{4a^2}\E .
	\end{array}
	\label{eq:n36-c4-boundaries}
\end{equation}
On the positive ray, a nonzero defect grows faster than either semisimple
contribution. On the negative ray, it grows at the same rate as the
$-a$ sector and survives together with its projector. In particular,
\begin{equation}
	b_{4,-}^2=P_{4,-},\qquad b_{4,-}^3=b_{4,-}^2.
	\label{eq:n36-c4-mixed}
\end{equation}
For $C_4\neq0$, this constant braid has rank three and Jordan form
$J_1(1)\oplus J_1(1)\oplus J_2(0)$. Thus no nonzero scalar multiple is
idempotent. For $C_4=0$, the nilpotent block vanishes and the limit is the
rank two projector $P_{4,-}$.

There is an important restriction on extending this observation to other
scalings. In the canonical coordinates of \eqref{eq:n36-c4-normal}, the
whole family $P_{4,-}-q\E$ obeys the constant braid relation. The analogous
family $P_{4,+}+q\E$ does not when $q\neq0$. With the braid defect defined
as $B_{12}B_{23}B_{12}-B_{23}B_{12}B_{23}$, its matrix element from
$\ket{011}$ to $\ket{000}$ is $-q$. Thus convergence along a joint
parameter and rapidity path must be distinguished from a verified constant
braid relation. 

\subsection{Exact matching observables}

The previous similarities contain $1/a$., to study $a\to0$, we should return instead to the original density and $R$-matrix and hold the other couplings
fixed. The corresponding limit is
\begin{equation}
	\begin{gathered}
		\nu_4=h_4\big|_{a=0},\qquad
		\nu_4^2=2a_2a_4\E,\qquad \nu_4^3=0,\\
		\check R_4(u)\longrightarrow
		e^{u\nu_4}=\id+u\nu_4+\tfrac12u^2\nu_4^2.
	\end{gathered}
	\label{eq:n36-c4-contracted}
\end{equation}
On the locus $a_2=a_4=\beta$, $a_3=\kappa$, the conditioned single
particle terms combine into unconditioned lowering operators. With
$\mu=2\beta$, the periodic sum and its exponential are
\begin{equation}
	\begin{aligned}
		\mathcal A_L&=\mu\sum_j e_j+\kappa\sum_j e_je_{j+1},\\
		e^{t\mathcal A_L}
		&=\prod_j(\id+\mu t e_j)
		\prod_j(\id+\kappa t e_je_{j+1}).
	\end{aligned}
	\label{eq:n36-matching-factorization}
\end{equation}
All monomials in the generator commute and each has square zero. The
factorization is therefore exact and no time discretization.

A nonzero term in this product cannot contain two selected bond factors
sharing a site, since their product contains $e_j^2$. The selected bonds
are consequently a matching of the cycle graph $C_L$ underlying the
periodic chain. Once the matching is chosen, an uncovered site can either
remain unchanged or be removed by its single site factor. Its combined
weight is given by $y=1+\mu t$, while each selected bond contributes $x=\kappa t$.
The flat covector $\bra\Omega=(\bra0+\bra1)^{\otimes L}$ sums all final
configurations and provides 
\begin{align}
	\bra\Omega e^{t\mathcal A_L}\ket{1^L}
	&=Z_{C_L}(y,x),\qquad y=1+\mu t,\quad x=\kappa t,\nonumber\\
	Z_{C_L}(y,x)
	&=\sum_{m=0}^{\lfloor L/2\rfloor}
	\frac{L}{L-m}\binom{L-m}{m}x^m y^{L-2m}.
	\label{eq:n36-matching-polynomial}
\end{align}
The coefficient counts matchings with $m$ bonds in the labelled cycle.
This is the standard monomer and dimer partition polynomial
\cite{HeilmannLieb1972}, realized here as an amplitude of the limiting
integrable chain. The exclusion of overlapping dimers is combinatorial,
not particle hopping, and $y$ is an amplitude weight rather than a particle
density.

The factorisation also determines the exact nilpotency index
$\nu(B)=\min\{p\geq1\mid B^p=0\}$,
\begin{equation}
	\nu(\mathcal A_L)=
	\begin{cases}
		L+1,&\mu\neq0,\\
		\lfloor L/2\rfloor+1,&\mu=0,\ \kappa\neq0,\\
		1,&\mu=\kappa=0.
	\end{cases}
\end{equation}
Indeed, occupation counting gives $\mathcal A_L^{L+1}=0$, while
$\bra{0^L}\mathcal A_L^L\ket{1^L}=L!\mu^L$ defines sharpness
for $\mu\neq0$. With only pair removal, a nonzero product selects
disjoint bonds and the maximum matching size $\lfloor L/2\rfloor$
gives the second index.

For $\mu,\kappa\geq0$, each occupied site contributes $\mu$ and each
occupied bond contributes $\kappa$ to the column sum. Its maximum
$L(\mu+\kappa)$ is obtained on $\ket{1^L}$, hence 
\begin{equation}
	Q_{\rm md}=\mathcal A_L-L(\mu+\kappa)\id
	\label{eq:n36-matching-generator}
\end{equation}
generates a Markov process with killing. Independent clocks of rate
$\mu$ at each site and $\kappa$ at each bond remove an occupied site
or an occupied pair respectively or kill the process otherwise.
The scalar shift preserves all off diagonal amplitudes and converts
their flat sum into the survival probability
\begin{equation}
	\mathbb P_{1^L}(\text{survival to }t)
	=e^{-L(\mu+\kappa)t}Z_{C_L}(1+\mu t,\kappa t).
	\label{eq:n36-matching-survival}
\end{equation}

Conservative completion instead replaces the scalar subtraction by
configuration dependent escape terms. Consider the density
\begin{equation}
	M(\alpha,\beta,\kappa)=
	\begin{pmatrix}
		0&\alpha&\alpha&\kappa\\
		0&-\alpha&0&\beta\\
		0&0&-\alpha&\beta\\
		0&0&0&-2\beta-\kappa
	\end{pmatrix},
	\label{eq:n36-completion-death}
\end{equation}
where $\alpha$ is the deletion rate on a mixed bond, $\beta$ is
each single particle deletion rate from $11$, and $\kappa$ is its
pair deletion rate. The columns sum to zero identically, but for
nonnegative rates an analytic Baxterisation of difference form
with $\check R(0)=\id$ and $\check R'(0)=M$ exists precisely when
\begin{equation}
	\beta=\kappa=0
	\qquad\text{or}\qquad
	\alpha=\beta+\kappa.
	\label{eq:n36-completion-death-branches}
\end{equation} 
The symmetric erosion kernel
establishes sufficiency on the first branch. On the second, the
projector $\rho_0=\bar n+e$ resets either local basis state to
$\ket0$. So when $\ell=\rho_0-\id=e-n$ one obtains
\begin{equation}
	M=\beta(\ell\otimes\id+\id\otimes\ell)
	+\kappa(\rho_0\otimes\rho_0-\id).
	\label{eq:n36-reset-density}
\end{equation}
Overlapping densities commute because they are polynomials in
commuting single site resets, so $\check R(u)=e^{uM}$ supplies
the required Baxterisation. The dynamics combines independent
single particle deaths with bond shocks resetting both endpoints,
a nearest neighbor specialization of the construction of Marshall
and Olkin \cite{MarshallOlkin1967}.

On the matching locus $\alpha=\beta$, any $\kappa>0$ violates both
conditions in \eqref{eq:n36-completion-death-branches}. Thus the
matching amplitude and its killed interpretation remain valid,
but this particular conservative completion does not admit Baxterisation. Class~6 will use both this distinction and
the pure pair specialization at` $\mu=0$ of
\eqref{eq:n36-matching-survival}.

\section{Class 5: Degenerations with parity controlled deletion}
\label{sec:n36-c5}

Classes~5 and~6 retain the XXX exchange term with a lowering deformation. Exchange preserves occupation, whereas the additional terms
remove particles and therefore do not represent asymmetric hopping. For
Class~5, the limit in which exchange vanishes leads to endpoint deletion
with opposite signs at the two ends of a cluster. The distinction between
even and odd periodic chains determines whether a diagonal change of basis
can make these amplitudes nonnegative.

\subsection{Exchanges and boundary algebra}

Let us set $a=a_1$, $b=a_2$ and $c=a_3$ within this section. Removing the scalar
factor $1-au$ from the initial braid matrix gives
\begin{equation}
	\begin{aligned}
		h_5&=a(2P-\id)+\mathfrak n_5,\\
		\mathfrak n_5&=\frac{b+c}{2}(z\otimes e-e\otimes z)
		+\frac{b-c}{2}(\id\otimes e-e\otimes\id),\\
		\mathcal R_5(u)&=\id+2auP+u\mathfrak n_5+bc\,u^2\E,\qquad
		\check R_5(u)=(1-au)\mathcal R_5(u).
	\end{aligned}
	\label{eq:n36-c5-input}
\end{equation}
Both normalizations are regular at zero, but
$\mathcal R_5'(0)=h_5+a\id$ and $\check R_5'(0)=h_5$.
The nilpotent operator satisfies
$\mathfrak n_5^2=2bc\E$, $\mathfrak n_5^3=0$ and
$\{P,\mathfrak n_5\}=0$. This anticommutation distinguishes the local
algebra from Class~6, where the deformation commutes with the permutation.

The term proportional to $b-c$ is a lattice coboundary. Summing the density
on a periodic chain leaves
\begin{equation}
	\mathcal H_5^{(L)}=a\sum_j(2P_{j,j+1}-\id)+V_5,\qquad
	V_5=\chi\sum_j e_j(n_{j+1}-n_{j-1}),\qquad \chi=b+c.
	\label{eq:n36-c5-global}
\end{equation}
The periodic parameter reduction and the Jordan structure of the transfer
matrix are studied in \cite{NietoGarcia2024}. A basis transformation
depending on the two rapidities also removes one deformation parameter
from the local spectral representative \cite{deLeeuwFontanellaNieto2026}.
For the homogeneous periodic transfer matrix, the induced auxiliary
similarity cancels under the trace. Thus varying $b-c$ at fixed $b+c$
need not change the periodic hierarchy. This redundancy is distinct from
the limit $a\to0$, which removes the exchange term itself.

At $\chi=0$, the homogeneous periodic hierarchy has an XXX case.
Subtracting $aL\id$ from its Hamiltonian gives the symmetric simple
exclusion process (SSEP), with hopping rate $D=2a\geq0$.
For $D>0$, its local spectral kernel and time semigroup obey
\begin{equation}
	K_{\rm SEP}(u)=\frac{\id+DuP}{1+Du},\qquad
	e^{tD(P-\id)}
	=K_{\rm SEP}\!\left(\frac{\tanh(Dt)}{D}\right).
	\label{eq:n36-sep}
\end{equation}
The rational kernel follows from the standard Baxterisation of the permutation operator $P$, while the time relation follows from $P^2=\id$. The kernel is stochastic for every
$u\geq0$, but $t\geq0$ reaches only $0\leq Du<1$. Consequently, the
limit $t\to\infty$ gives $(\id+P)/2$, whereas $u\to\infty$ gives the
deterministic swap $P$ and for $D=0$ both families are the identity. These
local statements refer to the undeformed spectral representative, not to
a density with $b+c=0$ before the rapidity dependent basis transformation.

For $a\neq0$, the deformed local algebra also separates a finite fusion
point from the large rapidity limits. Define
\begin{equation}
	Q_5=\frac{\id-P}{2}-\frac{\mathfrak n_5}{4a}
	+\frac{bc}{8a^2}\E,\qquad
	T_5=\id-2Q_5,\qquad J_5=\frac{bc}{2a}\E.
	\label{eq:n36-c5-boundary-operators}
\end{equation}
Then
\begin{equation}
	\mathcal R_5(u)=(\id+2auT_5)(\id+uJ_5),\qquad
	Q_5^2=Q_5,\quad \rank Q_5=1,\quad T_5^2=\id.
	\label{eq:n36-c5-factorization}
\end{equation}
Indeed, $J_5^2=0$ and $T_5J_5=J_5T_5=J_5$. Expanding the product,
the linear terms proportional to $\E$ cancel and the quadratic term is
$bc\,u^2\E$. At $u=-1/(2a)$, the factorization gives
$\mathcal R_5(u)/2=Q_5$.

Although $Q_5$ is a rank one fusion projector, it does not satisfy the
constant braid equation. Its three site relations appear in the following form 
\begin{equation}
	Q_{5,12}Q_{5,23}Q_{5,12}=\tfrac14Q_{5,12},\qquad
	Q_{5,23}Q_{5,12}Q_{5,23}=\tfrac14Q_{5,23}.
	\label{eq:n36-c5-fusion-relations}
\end{equation}
The right hand sides act on different bonds. By contrast, substituting
$T_5=\id-2Q_5$ into the braid defect and using these identities gives zero.
Thus $T_5$ is an involutive constant braid. At fixed couplings, the leading
polynomial degree gives
\begin{equation}
	\begin{aligned}
		u^{-2}\mathcal R_5(u)&\longrightarrow bc\E &&(bc\neq0),\\
		u^{-1}\mathcal R_5(u)&\longrightarrow2aP+\mathfrak n_5=2aT_5
		&&(bc=0,\ a\neq0).
	\end{aligned}
	\label{eq:n36-c5-constants}
\end{equation}
The generic limit is square zero. When the quadratic coefficient vanishes,
the surviving braid is invertible for $a\neq0$. A simultaneous scaling
that retains both the permutation and pair term will be treated together
with the corresponding Class~6 limit.

\subsection{Parity dependence and protected monomers}

Taking $a\to0$ at fixed $b,c$ gives
$\mathcal R_5(u)=e^{u\mathfrak n_5}$ and periodic Hamiltonian $V_5$.
The bound $V_5^{L-1}=0$ is established in
\cite{NietoGarcia2024}, Section~5. We now determine whether the signed
matrix elements in this limit can be made nonnegative by a similarity
that is diagonal in the configuration basis.

Assume $\chi>0$. A particle can be deleted only when its two neighboring
occupations differ. For a cluster bounded by vacancies, removal of its left
endpoint has amplitude $+\chi$ and removal of its right endpoint has
amplitude $-\chi$. A singleton and the fully occupied periodic chain
are annihilated by $V_5$ and every permitted deletion preserves the number of
occupied clusters.

	\label{prop:n36-c5-gauge}
	For even $L$, the diagonal involution can be obtained in the form 
	\begin{equation}
		D_5\ket{\boldsymbol\eta}
		=(-1)^{\sum_{j=1}^L j\eta_j(1-\eta_{j+1})}
		\ket{\boldsymbol\eta},\qquad \eta_{L+1}=\eta_1,
		\label{eq:n36-c5-phase}
	\end{equation}
	which transforms $V_5$ into 
	\begin{equation}
		W_5=D_5V_5D_5^{-1}
		=\chi\sum_j e_j
		(\bar n_{j-1}n_{j+1}+n_{j-1}\bar n_{j+1}),
		\label{eq:n36-c5-positive}
	\end{equation}
	whose entries are nonnegative. For odd $L$, no invertible diagonal
	similarity in the configuration basis, including one with complex
	diagonal entries, makes all nonzero matrix elements of $V_5$
	nonnegative real numbers.

	The exponent in \eqref{eq:n36-c5-phase} is the sum of the labelled
	positions of the right endpoints of occupied clusters, modulo two.
	Removing a left endpoint leaves this sum unchanged. Removing a right
	endpoint moves it from $j$ to $j-1$ and changes its parity, cancelling
	the negative amplitude. Across the bond connecting sites $L$ and $1$,
	the corresponding shift is from $1$ to $L$. Its parity changes exactly
	when $L$ is even. The displayed transformation therefore works for
	every allowed deletion on an even periodic chain.
	
	We can prove the obstruction, by letting $d_j$ be the nonzero diagonal entry
	assigned to the singleton at $j$ and let $d_{j,j+1}$ be the entry
	assigned to the cluster occupying $j,j+1$. The two deletions of that cluster have opposite amplitudes. Positivity of both transformed
	entries forces $d_j/d_{j+1}$ to be a negative real number.
	Multiplying these ratios around the periodic chain must give one,
	which is impossible for odd $L$. Here so far we make no assumption that the initial diagonal entries are real.

Under $W_5$, each occupied cluster bounded by vacancies loses its endpoints
until one protected particle remains. In a nonuniform configuration, call
such a cluster active when its length is at least two, and let $k_{\rm act}$
be their number. Each contributes two amplitudes $\chi$, so the column sum
is $2\chi k_{\rm act}$. An active cluster requires at least two occupied
sites and a separating vacancy. Thus
$k_{\rm act}\leq\lfloor L/3\rfloor$, with equality attainable.
The fully occupied configuration has no endpoints and contributes zero.
Consequently,
\begin{equation}
	\mathcal Q_5^{(L)}=W_5-\lambda\id,\qquad
	\lambda\geq2\chi\lfloor L/3\rfloor
	\label{eq:n36-c5-killed}
\end{equation}
defines a killed Markov generator on an even periodic chain. Its killing
rate in a configuration is $\lambda-2\chi k_{\rm act}$. Protected
monomers are inert under deletion, but still subject to this killing.

The transformation $D_5$ acts on the full configuration space. It is not a
tensor product of single site gauges or a positive Doob transform.
Conjugating a commuting hierarchy by $D_5$ preserves commutation, but the
identity \eqref{eq:n36-c5-positive} does not construct a stochastic two
site $R$ matrix for $W_5$. 

\subsection{Exact finite time amplitudes}

Let a nonuniform initial configuration have occupied clusters of lengths
$m_1,\ldots,m_k$, separated by vacancies. These vacancies remain empty,
so the initial clusters cannot merge. A cluster of length $m$ permits
$j$ successive endpoint removals for $0\leq j\leq m-1$. There are two
choices at each step, and their combined amplitude is $2\chi$. The sum
over final configurations after exactly $j$ removals is therefore
$(2\chi)^j$. Including the exponential ordering factor gives
\begin{equation}
	F_m(t)=\sum_{j=0}^{m-1}\frac{(2\chi t)^j}{j!}.
	\label{eq:n36-c5-cluster-amplitude}
\end{equation}
For distinct initial clusters, interleaving the ordered histories supplies
the multinomial factors that multiply these exponential generating
amplitudes. With the flat covector defined in Class~4,
\begin{align}
	\bra\Omega e^{tW_5}\ket{\boldsymbol\eta}
	&=\prod_{i=1}^k F_{m_i}(t),\nonumber\\
	\mathbb P_{\boldsymbol\eta}(\text{survival to }t)
	&=e^{-\lambda t}\prod_{i=1}^k
	\sum_{j=0}^{m_i-1}\frac{(2\chi t)^j}{j!}.
	\label{eq:n36-c5-survival}
\end{align}
The first line is a non-normalized amplitude and the second is the survival
probability for \eqref{eq:n36-c5-killed}. Here we have only one factor
$e^{-\lambda t}$, because the scalar shift acts on the entire chain.

A single cluster of length $L-1$ attains time degree $L-2$, which together with
the upper bound and the even length similarity, this gives nilpotency
index $L-1$ for the positive representative and for $V_5$ when $L$ is
even. The fully occupied configuration has no permitted deletion and its
survival probability is $e^{-\lambda t}$. The empty configuration obeys
the same law, represented by the empty product in
\eqref{eq:n36-c5-survival}. The formulas also describe the operator $W_5$ defined by its positive
transition rules on a periodic chain of odd length. In that case,
Proposition~\ref{prop:n36-c5-gauge} excludes its identification with
$V_5$ by a diagonal similarity. The counting identity itself is independent
of this qualification, but its interpretation as a transformed Class~5
model is not.

\section{Class 6: Fragmentation, parity and the common pair boundary}
\label{sec:n36-c6}

In Class~6, the lowering deformation is symmetric under interchange of the
two sites. The generic periodic model has a Jordan obstruction at the
spectral value that would be stationary in a conservative Markov process.
After the exchange term is removed, however, a phase transformation makes
the remaining deletion amplitudes nonnegative for every chain length.
The parity dependence then concerns the existence of complete deletion
histories, rather than the existence of a positive representative.

\subsection{Exchange and obstruction at the stationary eigenvalue}

Set $a=a_1$ and $b=a_2$. With the polynomial normalization already used for Class~5, the local density and $R$-matrix are
\begin{equation}
	\begin{aligned}
		h_6&=a(2P-\id)+\mathfrak n_6,\qquad
		\mathfrak n_6=b(z\otimes e+e\otimes z),\\
		\mathcal R_6(u)&=\id+2auP+u(1+2au)\mathfrak n_6
		-b^2u^2(1+2au)^2\E.
	\end{aligned}
	\label{eq:n36-c6-input}
\end{equation}
Here $\check R_6(u)=(1-au)\mathcal R_6(u)$ and
$\mathcal R_6'(0)=h_6+a\id$. The local relations are
\begin{equation}
	[P,\mathfrak n_6]=0,\qquad
	\mathfrak n_6^2=-2b^2\E,\qquad \mathfrak n_6^3=0.
	\label{eq:n36-c6-local-algebra}
\end{equation}
Thus permutation preserves the lowering operator rather than reversing
its sign as in Class~5. The relation to the rational eleven vertex model
is established in \cite{AtalikovZotov2023}. For the present limits, the
relevant distinction is the commutation in
\eqref{eq:n36-c6-local-algebra}. The periodic Hamiltonian is
\begin{equation}
	\mathcal H_6^{(L)}=a\sum_j(2P_{j,j+1}-\id)+V_6,\qquad
	V_6=b\sum_j e_j(z_{j-1}+z_{j+1}).
	\label{eq:n36-c6-global}
\end{equation}
Exchange preserves occupation, while $V_6$ lowers it by one. Ordering the
basis by occupation therefore gives a block triangular matrix whose
diagonal blocks are those of the exchange Hamiltonian. This determines
the eigenvalues, but not the Jordan structure.

For $a>0$, put $A_6=\mathcal H_6^{(L)}-aL\id$. Its diagonal blocks are
those of the symmetric exclusion generator
$2a\sum_j(P_{j,j+1}-\id)$. Hence all eigenvalues are real and
nonpositive, and their largest real part is zero. Let
$\ket W=\sum_j\ket{0\cdots1_j\cdots0}$. Exchange annihilates both
$\ket W$ and the vacuum. Each singleton is removed with amplitude $2b$,
so
\begin{equation}
	A_6\ket{0^L}=0,\qquad A_6\ket W=2bL\ket{0^L}.
	\label{eq:n36-c6-zero-jordan}
\end{equation}
For $b\neq0$, this is a nontrivial Jordan chain at zero. Any scalar shift
leading to a conservative finite Markov generator must keep its largest
real eigenvalue at zero. The shift is consequently fixed, and an invertible
similarity cannot remove the chain. Its exponential grows linearly on
$\ket W$, contradicting boundedness of a finite stochastic semigroup.
Therefore no global similarity and scalar shift turn the stated periodic
operator into a conservative Markov generator. This argument is stronger
than the failure of a particular diagonal gauge. It does not exclude a
killed process or a singular parameter limit.

\subsection{Exchange free fragmentation}

At $a=0$, the local spectral matrix is $e^{u\mathfrak n_6}$ and the
periodic Hamiltonian is $V_6$. Put $r=2b$. For an occupied site $j$ in a
configuration $\boldsymbol\eta$, the deletion amplitude is
$r(1-\eta_{j-1}-\eta_{j+1})$. An isolated particle has amplitude $+r$,
a particle with two occupied neighbors has amplitude $-r$, and a particle
whose neighboring occupations differ cannot be deleted. An interior
deletion splits a cluster into two nonempty intervals. A dimer bounded by
vacancies has neither a removable interior nor an isolated particle and
is therefore blocked.

Let $f_m(t)$ be the amplitude for completely emptying an occupied interval
of length $m$ bounded by vacancies. Every complete history contains exactly
$m$ deletions. For a singleton, $f_1(t)=rt$. For $m\geq2$, the first
allowed deletion lies in the interior and leaves intervals of lengths
$j$ and $m-1-j$. They subsequently evolve independently and integrating over
the first deletion time and differentiating gives
\begin{equation}
	\begin{gathered}
		f_1(t)=rt,\qquad f_m(0)=0,\\
		f_m'(t)=-r\sum_{j=1}^{m-2}f_j(t)f_{m-1-j}(t)\qquad(m\geq2).
	\end{gathered}
	\label{eq:n36-c6-emptying-recursion}
\end{equation}
For $m=2$ the sum is empty. The minus sign is the amplitude of the first
interior deletion. The product already includes the interleaving of later
events, since the $f_j$ are exponential generating amplitudes in time.

The formal series $F(\zeta,t)=\sum_{m\geq1}f_m(t)\zeta^m$ satisfies
$\partial_tF=r\zeta(1-F^2)$ and $F(\zeta,0)=0$. Therefore
\begin{equation}
	\sum_{m\geq1}f_m(t)\zeta^m=\tanh(rt\zeta).
	\label{eq:n36-c6-tanh}
\end{equation}
The even coefficients vanish because no complete history exists at those
lengths. Indeed, if a complete history has $I$ interior deletions and $S$
singleton deletions, its cluster count and total number of removals give
$1+I-S=0$ and $m=I+S$. Thus $m=2I+1$ must be odd. For an odd length,
all complete histories have the same sign $(-1)^I$. The alternating signs
in the series record this fact, not a cancellation between complete
histories with different signs.

	\label{prop:n36-c6-index}
	For $b\neq0$ and $L\geq3$, the periodic generator in the limit $a=0$
	has nilpotency index
	\begin{equation}
		\nu(V_6)=
		\begin{cases}
			L+1,&L\text{ even},\\
			L-1,&L\text{ odd}.
		\end{cases}
		\label{eq:n36-c6-index}
	\end{equation}
	Every nonzero action removes one particle $V_6^{L+1}=0$.
	Write $c_m=[x^m]\tanh x$. Starting from the fully occupied periodic
	chain, the first deletion has $L$ choices, each with amplitude $-r$.
	The remaining interval has length $L-1$ and is bounded by the new
	vacancy. Hence
	\begin{equation}
		\bra{0^L}e^{tV_6}\ket{1^L}
		=-Lr\int_0^t f_{L-1}(s)\,ds
		=-c_{L-1}(rt)^L.
		\label{eq:n36-c6-full-empty}
	\end{equation}
	For even $L$, $c_{L-1}\neq0$, so $V_6^L\neq0$ and the index is $L+1$.
	
	For odd $L$, an entry of $V_6^{L-1}$ could only connect occupation
	$L-1$ to zero or occupation $L$ to one. The first case starts from a
	single interval of even length $L-1$, for which complete emptying
	is impossible. To treat the second, exchange occupied and vacant
	states with $\mathcal F=(\sigma^x)^{\otimes L}$. Since
	$\mathcal F e_j\mathcal F=e_j^{\mathsf T}$ and
	$\mathcal F z_j\mathcal F=-z_j$, one has
	$\mathcal F V_6\mathcal F=-V_6^{\mathsf T}$. This relates every
	matrix element of the second type, up to a sign, to one of the first
	type. Thus $V_6^{L-1}=0$. Finally, an occupied interval of odd length
	$L-2$ with two vacant sites outside it has a nonzero emptying
	amplitude. Therefore $V_6^{L-2}\neq0$ and the index is $L-1$.

This dependence on $L$ is a property of the longest admissible global
histories. The local operator $\mathfrak n_6$ has nilpotency index three
for every $b\neq0$. Unlike the sign obstruction in Class~5, the change in
global index does not prevent a positive phase representative.

\subsection{Positive phase representative}

Assign a phase to each configuration according to its number of occupied
nearest neighbor bonds 
\begin{equation}
	D_6\ket{\boldsymbol\eta}
	=i^{\sum_j\eta_j\eta_{j+1}}\ket{\boldsymbol\eta},\qquad
	D_6=\exp\!\left(\frac{i\pi}{2}\sum_jn_jn_{j+1}\right).
	\label{eq:n36-c6-phase}
\end{equation}
An isolated deletion does not change an occupied bond. An interior deletion removes
two such bonds and multiplies its amplitude by $i^{-2}=-1$. Thus, for
$r>0$,
\begin{equation}
	W_6=D_6V_6D_6^{-1}
	=r\sum_j e_j(\bar n_{j-1}\bar n_{j+1}+n_{j-1}n_{j+1})
	\label{eq:n36-c6-positive}
\end{equation}
has nonnegative entries. The change in the occupied bond count is the same
across the periodic boundary, so the transformation works for both even
and odd $L$. Although $D_6$ is a product of commuting bond factors, it is
not a tensor product of single site gauges. As for $D_5$, it is a global
phase similarity rather than a positive Doob transform or a local stochastic
normalization of the original $R$ matrix.

The column sum of $W_6$ is $r$ times the number of occupied sites with
equal neighboring occupations. It is at most $rL$, attained in the fully
occupied configuration. Therefore
\begin{equation}
	\mathcal Q_6^{(L)}=W_6-\sigma\id,\qquad \sigma\geq rL
	\label{eq:n36-c6-killed-generator}
\end{equation}
is again a killed Markov generator. At $\sigma=rL$, give each site a clock of
rate $r$. An allowed event removes that particle, while any other event
kills the process. A larger $\sigma$ adds uniform killing. Separated
occupied dimers are inert under $W_6$, but remain subject to this loss of
probability.

To calculate survival, one must retain all final configurations rather
than only the vacuum. Let $Z_m(x)$ be the flat amplitude starting from a
vacancy bounded occupied interval of length $m$, with dimensionless time
$x=rt$. The history with no deletion gives $Z_m(0)=1$. A singleton can
remain or disappear, while a dimer cannot change. Decomposing the longer
intervals according to their first deletion gives
\begin{equation}
	Z_1(x)=1+x,\qquad Z_2(x)=1,\qquad
	Z_m'(x)=\sum_{j=1}^{m-2}Z_j(x)Z_{m-1-j}(x)\quad(m\geq2).
	\label{eq:n36-c6-partial-recursion}
\end{equation}
Compared with \eqref{eq:n36-c6-emptying-recursion}, the interior amplitude
is now positive and either daughter interval may remain partially occupied.
These changes account for the sign and the nonzero initial data. For $\mathcal Z(\zeta,x)=\sum_{m\geq1}Z_m(x)\zeta^m$, the recursion is
\begin{equation}
	\partial_x\mathcal Z=\zeta(1+\mathcal Z^2),\qquad
	\mathcal Z(\zeta,0)=\frac{\zeta}{1-\zeta}.
	\label{eq:n36-c6-partial-ode}
\end{equation}
Solving the Riccati equation gives
\begin{equation} 
	\displaystyle
		\mathcal Z(\zeta,x)=
		\frac{\zeta+(1-\zeta)\tan(x\zeta)}
		{1-\zeta-\zeta\tan(x\zeta)}.
	\label{eq:n36-c6-partial-series}
\end{equation}
For instance, $Z_3(x)=1+x+x^2+x^3/3$ and $Z_4(x)=(1+x)^2$.
The equation is understood as a formal generating series in $\zeta$.
Each coefficient is a finite polynomial in time, so poles of the summed
function do not indicate a finite time singularity in any finite chain.

For a nonuniform initial configuration with occupied cluster lengths $m_i$,
the original vacancies permanently separate those clusters. Their histories
interleave as in Class~5, and the flat amplitude is $\prod_iZ_{m_i}(rt)$.
The single global killing factor then gives
\begin{equation}
	\mathbb P_{\boldsymbol\eta}(\text{survival to }t)
	=e^{-\sigma t}\prod_i Z_{m_i}(rt).
	\label{eq:n36-c6-survival}
\end{equation}
The fully occupied periodic chain requires its own first deletion. There
are $L$ possible positions, after which the occupied interval has length
$L-1$. Its flat amplitude is consequently
\begin{equation}
	1+L\int_0^{rt}Z_{L-1}(x)\,dx.
	\label{eq:n36-c6-full-partial}
\end{equation}
This replaces the product in \eqref{eq:n36-c6-survival}. The empty
configuration is represented by the empty product and has survival
$e^{-\sigma t}$. Together these formulas cover every initial configuration.

For complete deletion alone, the initial generating series is zero and its
positive solution is $\tan(x\zeta)$. So by considering $\tan x=\sum_{m\geq1}\mathsf T_mx^m/m!$, the integer $\mathsf T_m$
counts the permitted ordered complete histories of an interval of length
$m$. Every such history carries the same positive weight, and the
exponential contributes the factor $1/m!$. This identifies the counts
with tangent numbers and again excludes even lengths. The comparison
with \eqref{eq:n36-c5-survival} is therefore structural. Class~5 erodes
endpoints and preserves a monomer in each initial cluster. Class~6 splits
cluster interiors, can remove singletons, and leaves dimers blocked.

\subsection{Common permutation pair boundary}

In the previous limit held rapidity fixed and at  fixed couplings large rapidity instead selects the leading polynomial degree 
\begin{equation}
	\begin{aligned}
		u^{-4}\mathcal R_6(u)&\longrightarrow-4a^2b^2\E &&(ab\neq0),\\
		u^{-2}\mathcal R_6(u)&\longrightarrow-b^2\E &&(a=0,\ b\neq0),\\
		u^{-1}\mathcal R_6(u)&\longrightarrow2aP &&(b=0,\ a\neq0).
	\end{aligned}
	\label{eq:n36-c6-leading-limits}
\end{equation}
For $a=b=0$ the polynomial spectral family is the identity. To retain
permutation together with a nonzero pair term, let $u=U/\varepsilon$ with
fixed $a\neq0$, $U\neq0$, and take $\varepsilon\to0^+$. The two
exchange classes then admit the respective scalings as follows 
\begin{align}
	\frac{\varepsilon}{2aU}\mathcal R_5(U/\varepsilon)
	&\longrightarrow P+\frac{\beta\gamma}{2a}U\E,
	&& b=\beta\sqrt\varepsilon,\quad c=\gamma\sqrt\varepsilon,
	\nonumber\\
	\frac{\varepsilon}{2aU}\mathcal R_6(U/\varepsilon)
	&\longrightarrow P-2a\eta^2U^3\E,
	&& b=\eta\varepsilon^{3/2}.
	\label{eq:n36-common-double-scaling}
\end{align}
For Class~5, $bc=O(\varepsilon)$ balances its quadratic term against the
linear permutation term. For Class~6, $b^2=O(\varepsilon^3)$ balances its
quartic term against the same scale. The single particle lowering terms
vanish after normalization in both cases. The different powers of $U$
therefore encode the polynomial degrees of the two source families. For a fixed coefficient both paths reach
\begin{equation}
	B_\kappa=P+\kappa\E,\qquad
	B_\kappa^{-1}=P-\kappa\E,\qquad
	B_\kappa^2=\id+2\kappa\E.
	\label{eq:n36-common-braid}
\end{equation}
Its constant braid relation has an independent proof. The unbraided matrix
is $PB_\kappa=\id+\kappa e\otimes e$. Its three embeddings are
polynomials in commuting single site operators $e_j$, so the constant
quantum Yang--Baxter equation holds. Multiplication by the permutation
then gives the braid equation. The inverse and square follow from
$P\E=\E P=\E$ and $\E^2=0$.

Each fixed $U$ therefore gives a constant braid and the limiting families in
$U$ are not regular at zero, since their continuations equal $P$ rather
than $\id$. The normalization in
\eqref{eq:n36-common-double-scaling} also excludes $U=0$, this is distinct from the regular spectral limit constructed below.

For $\kappa\geq0$, the scalar normalization $B_\kappa/(1+\kappa)$ is
substochastic and preserves the braid equation. Its column sums are
$(1,1,1,1+\kappa)/(1+\kappa)$. Normalizing the columns separately does
not have the same algebraic property. With
$\widehat B=B_\kappa\operatorname{diag}(1,1,1,q)$ and
$q=(1+\kappa)^{-1}$, the matrix element from $\ket{111}$ to
$\ket{001}$ of its braid defect is $-(q-1)^2(q+1)$.
For $\kappa>0$ this is nonzero. Positivity of the pair augmented swap
therefore does not make its column normalized version a conservative
integrable exclusion update.

Conventionally one starts instead from a singular product similarity. Take $g_s=\id+se$ with $s=-\kappa/(4b)$ and then let $b\to0$ at fixed
$a$, $\kappa$ and $u$. The product shear leaves $P$ and $\E$ unchanged,
while
\begin{equation}
	(g_s\otimes g_s)\mathfrak n_6(g_s\otimes g_s)^{-1}
	=\mathfrak n_6-4bs\E.
	\label{eq:n36-c6-pair-shear}
\end{equation}
Although the similarity itself diverges, the transformed density and $R$-matrix converge to
\begin{equation}
	h_{\rm pair}=a(2P-\id)+\kappa\E,\qquad
	\mathcal R_{\rm pair}(u)=(\id+2auP)(\id+\kappa u\E).
	\label{eq:n36-pair-regular}
\end{equation}
As before, $\mathcal R_{\rm pair}'(0)=h_{\rm pair}+a\id$, and the source
scalar factor restores derivative $h_{\rm pair}$. Taking $a\to0$ next
gives exactly the $\mu=0$ matching sector of Class~4. Its nilpotency index,
flat amplitude and killed survival probability follow from
\eqref{eq:n36-matching-factorization} and
\eqref{eq:n36-matching-survival} without another counting argument. Finally, adding the pair escape term gives a different conservative density 
\begin{equation}
	M_{\rm AD}(D,\lambda)=D(P-\id)+\lambda(\E-n\otimes n).
	\label{eq:n36-ad-completion}
\end{equation}
For $D,\lambda\geq0$, it describes symmetric hopping
$01\leftrightarrow10$ at rate $D$ and annihilation $11\to00$ at rate
$\lambda$. Column conservation holds throughout this quadrant. However for a regular analytic Baxterisation of difference form whose derivative is
exactly this density, the consistency test in
Appendix~\ref{app:n36-completions} requires $\lambda=0$ or $\lambda=2D$. The first case is \eqref{eq:n36-sep}, for the second we define
\begin{equation}
	\begin{aligned}
		B_{\rm A}&=P-2n\otimes n+2\E,\qquad B_{\rm A}^2=\id,\\
		K_{\rm AD}(u)&=\frac{\id+DuB_{\rm A}}{1+Du},\qquad 0\leq Du\leq1.
	\end{aligned}
	\label{eq:n36-ad-kernel}
\end{equation}
The constant braid relation for $B_{\rm A}$ gives the rational spectral
family, and its derivative is
$D(B_{\rm A}-\id)=M_{\rm AD}(D,2D)$. The stated positivity interval is
essential, since the $\ket{11}$ survival entry is $(1-Du)/(1+Du)$, which becomes
negative beyond $Du=1$. For $D>0$, the substitution
$u=\tanh(Dt)/D$ realizes the entire semigroup for $t\geq0$ within this
interval. For $D=0$, the surviving branch is the zero generator.

The nontrivial branch is the established free fermion annihilating random
walk, equivalently the domain wall process of Glauber dynamics at zero
temperature \cite{AlcarazDrozHenkelRittenberg1994,Glauber1963}.
It is obtained as a separately checked conservative completion, not as a
similarity representative of the generic Class~6 model. The parameter and
similarity limits leading to pair amplitudes preserve their indicated
spectral equations, whereas the modification of escape rates requires an independent integrability test.

\newpage 

\section{Discussion and remarks}
\label{sec:discussion}

We have related six regular integrable classes to their finite periodic dynamics through local operator algebras and distinct limits. Minimal polynomials determined nilpotency and the possible polynomial factors in time
evolution. We have found that spectral limits isolate idempotent, nilpotent and mixed constant braid operators. The analysis also distinguished local from global diagonalisability, spectral evolution processes and distinct probability conservation cases.

We have established that the erosion model determines maximal Jordan block sizes at every
eigenvalue and exhibits a change of Jordan partition at fixed spectrum.
The deletion limits distinguished the parity obstruction to removing signs by a diagonal similarity in Class 5 from parity dependent nilpotency in
Class 6. In the latter case we found that matching and fragmentation expansions give exact amplitudes and survival probabilities, while local
consistency tests provided the admissible conservative completions of the specified densities. These results concern finite chains, however it yet remains to determine the indecomposable structure of the full conserved hierarchy and dependence on boundary conditions.

Current progress includes the construction of quantum algebras defined by the quadratic exchange relations and determine their coproducts and indecomposable representations \cite{ReshetikhinTakhtadzhyanFaddeev1990}.
A cohomological classification should also allow to distinguish specific deformations of the
Yang-Baxter relations from central extensions
of the associated operator algebras. Computation of the cocycles and higher order obstructions would admit to test whether the defect parameters can describe
nontrivial extension classes or changes \cite{Eisermann2005}. 

A complementary direction that is also currently under active investigation is assembling the associated $R$-matrices into quantum circuit architectures and establish their conserved quantities through compatible inhomogeneous transfer matrices \cite{MiaoGritsevKurlov2024}. The generic operators are not unitary, so
physical realizations must distinguish stochastic updates from quantum operations. Their amplitudes could then be used to investigate
Lee-Yang zeros in complexified gate parameters and their evolution with
circuit depth \cite{LeeYang1952,JiangLiuWuZhang2026}. One of the questions is how nilpotent sectors and polynomial depth factors associated with Jordan blocks
modify the trajectories and accumulation sets of these zeros. 

Another important sector, includes the nilpotent and triangular structures that provide framework for seeking
new solutions with larger local spaces and higher rank quantum symmetries. For instance, given extensions could support coupled occupation and reaction processes with several species. Another possibility would be to address higher simplex relations \cite{Zamolodchikov1981,MailletNijhoff1989}. Here the goal would be to construct solutions leading to commuting layer transfer operators and novel dynamics on higher dimensional lattices. The specific boundary conditions and positivity constraining must be
established independently. This would also allow to further determine the mechanisms underlying the Jordan structure and stochastic limits that are beyond two local states and admit higher range interactions.

\section*{Acknowledgements}

The author thanks Konstantin Khanin, Nicolai Reshetikhin, Alexander Povolotsky, Pavel Pyatov, Dmitry Talalaev and Bart Vlaar for valuable discussions and useful comments. The research of AP was supported by the Beijing Natural Science Foundation Grant IS25017 and the Russian Science Foundation Grant RSCF-25-72-10177. The author is also grateful to Valery Gritsenko and Mikhail Alfimov for productive discussions and hospitality.

\newpage 
\appendix 
\section{Conservative completions}
\label{app:n36-completions}
For either density \eqref{eq:n36-completion-death} or
\eqref{eq:n36-ad-completion} a regular analytic Baxterisation of difference form with $\check R(0)=\id$ and $\check R'(0)=M$
requires the Reshetikhin condition
\cite[Eq.~(3.20)]{KulishSklyanin1982}
\begin{equation}
	C(M)=[M_{12}+M_{23},[M_{12},M_{23}]]=K_{23}-K_{12}
	\label{eq:n36-reshetikhin}
\end{equation}
for some two site operator $K$. The following functional
\begin{equation}
	\mathcal L(C)=\bra{000}C\ket{011}+\bra{001}C\ket{111}
	\label{eq:n36-completion-witness}
\end{equation}
annihilates $K_{23}-K_{12}$ because the two contributions are
$K_{00,11}$ and $-K_{00,11}$. Thus $\mathcal L(C(M))=0$ is necessary.

\paragraph{The deletion completion with three rates.}
For \eqref{eq:n36-completion-death}, direct substitution gives
\begin{equation}
	\mathcal L(C(M))=-2(\kappa+2\beta)(\alpha-\beta-\kappa)^2, 
	\label{eq:n36-death-obstruction}
\end{equation}
and for nonnegative rates its vanishing requires $\beta=\kappa=0$ or
$\alpha=\beta+\kappa$. Sufficiency follows from the
symmetric erosion kernel \eqref{eq:n36-c3-kernel} and the resetting representation \eqref{eq:n36-reset-density}, whose overlapping densities commute and admit $\check R(u)=e^{uM}$. 

\paragraph{The hopping and annihilation completion.}
For \eqref{eq:n36-ad-completion}, the same calculation yields
\begin{equation}
	\mathcal L(C(M_{\rm AD}))=2\lambda^2(2D-\lambda).
	\label{eq:n36-ad-obstruction}
\end{equation}
Hence $\lambda=0$ or $\lambda=2D$ is necessary. The kernels
\eqref{eq:n36-sep} and \eqref{eq:n36-ad-kernel} establish
sufficiency on the respective branches. These explicit constructions,
not the necessary condition alone, complete the argument.

\newpage 
\bibliographystyle{unsrt}
\bibliography{References}

\end{document}